\documentclass{JHEP3}
\usepackage{graphicx}
\usepackage{graphics}
\usepackage{cite}
\usepackage{epsfig}
\usepackage{amsmath}
\usepackage{mathrsfs}
\usepackage{amsthm}
\usepackage{amssymb}
\usepackage{oldgerm}
\usepackage{enumerate}
\usepackage{indentfirst}
\usepackage[latin1]{inputenc}
\usepackage{setspace}

\def\ss{\scriptscriptstyle}

\title{Quasinormal modes of plane-symmetric black holes according to the
AdS/CFT correspondence}

\author{Alex S. Miranda$^{\dag}$, Jaqueline
Morgan$^{\ddag,a}$, and Vilson T. Zanchin$^{\ddag,b}$    
 \\
$^{\dag}$Centro de Ci\^encias Naturais e Exatas,
Universidade Federal de Santa Maria,\\
97119-900 Santa Maria, RS, Brazil\\
email: amiranda@mail.ufsm.br\\
\\
$^\ddag$Centro de Ci\^encias Naturais e Humanas, Universidade Federal
do ABC,\\ Rua Santa Ad\'elia 166, 09210-170, Santo Andr\'e, SP, Brazil\\
$^{a}$email: jaqueline.morgan@ufabc.edu.br;
$^{b}$email: zanchin@ufabc.edu.br}

\abstract{The electromagnetic and gravitational quasinormal spectra of
$(3+1)$-dimensional plane-symmetric anti-de Sitter black holes are analyzed
in the context of the AdS/CFT correspondence. According to such a
correspondence, the electromagnetic and gravitational quasinormal frequencies
of these black holes are associated respectively to the poles of retarded
correlation functions of $R$-symmetry currents and stress-energy tensor in
the holographically dual conformal field theory: the $(2+1)$-dimensional
$\mathcal{N}=8$ super-Yang-Mills theory. The connection between AdS black
holes and the corresponding field theory is used to unambiguously fix the
boundary conditions that enter the proper definition of quasinormal modes.
Such a procedure also helps one to decide, among the various different
possibilities, what are the appropriate gauge-invariant quantities one should
use in order to correctly describe the electromagnetic and gravitational
blackhole perturbations. These choices imply in different dispersion
relations for the quasinormal modes when compared to some of the results in
the literature. In particular, the long-distance, low-frequency limit of
dispersion relations  presents the characteristic hydrodynamic behavior of a
conformal field theory with the presence of diffusion, shear, and sound wave
modes. There is also a family of purely damped electromagnetic modes which
tend to the bosonic Matsubara frequencies in the long-wavelength regime.}


\begin{document}

\section{Introduction}
\label{introduction}

\subsection{Motivations and overview}

Theoretical studies on black holes in asymptotically anti-de Sitter
spacetimes have attracted substantial attention since the advent of the
anti-de Sitter/conformal field theory (AdS/CFT) correspondence
\cite{maldacena1,gubser1,witten1}. In particular, the quasinormal-mode (QNM)
spectra of various types of asymptotically AdS black holes have been analyzed
since then (see Refs. \cite{wang1,govi1,wang2,zhu,wang3,konoplya1, konoplya2,
starinets1,aros,musiri1,musiri2,crisostomo,fernando,konoplya3,siopsis1,wang4,
giammatteo1,jing1,maeda,siopsis2,zhang,zhang2,zhidenko,rao,kout1,amado,aliev}
for a sample).
According to the AdS/CFT correspondence, an asymptotically AdS black hole is,
in the CFT side, associated to a system in thermal equilibrium whose
temperature is the Hawking temperature of the black hole. In such a context,
blackhole perturbations correspond to small deviations from equilibrium of
the CFT thermal system, and the characteristic damping time of perturbations,
which is given by the inverse of the imaginary part of the fundamental QNM
frequency, is a measure of the dynamical timescale of approach to thermal
equilibrium of the corresponding conformal field theory \cite{horowitz1}.

The literature on QNM of AdS black holes includes studies taking into account
a variety of different aspects such as the topology of the event horizon, the
number of dimensions of the spacetime, the particular type of perturbation
fields considered, and also the special parameters which characterize each
different black hole itself. Each one of these variant properties reflects on
the dual CFT. For instance, assuming the $(3+1)$-dimensional AdS spacetime
contains a plane-symmetric black hole, then the holographic field theory is
defined over the $(2+1)$-dimensional Minkowski spacetime, which is the
conformal boundary of the bulk AdS spacetime. Moreover, different blackhole
parameters characterize different dual plasmas in the CFT side, and different
equilibrium states of such systems at the boundary.

An important issue in the study of the vibrational modes of black holes is
the choice of appropriate boundary conditions. In the case of asymptotically
flat spacetimes, the solutions to the wave equations governing linear
perturbations are, near the boundaries, given by plane wave functions. QNM
are then defined as solutions which satisfy physically well motivated
boundary conditions, namely, purely ingoing waves at the horizon and purely
outgoing waves at infinity (see Refs. \cite{kokkotas, nollert} for reviews).
For anti-de Sitter black holes, on the other hand, the condition at the
future horizon is the same as for asymptotically flat spacetimes, but now
there are no natural conditions to be imposed on the perturbation variables
at the AdS infinity. These can be Dirichlet, Neumann, or Robin boundary
conditions, depending on whether it is required that the field perturbations,
their derivatives or a combination of both vanish at the AdS boundary,
respectively. In the study of the evolution of a massless scalar field in
$(3+1)$-, $(4+1)$-, and $(6+1)$-dimensional Schwarzschild-AdS spacetimes,
Horowitz and Hubeny \cite{horowitz1} computed the corresponding
quasinormal-mode
spectra by imposing Dirichlet boundary conditions on such a field at
infinity. This option was well justified in that context, since by writing
the radial part of the Klein-Gordon equation in a Schr\"odinger-like form,
the resulting effective potential diverges at that boundary. The same
boundary condition was used to study massless scalar and electromagnetic
perturbations of $(2+1)$-dimensional Ba\~{n}ados-Teitelboim-Zanelli (BTZ)
black holes \cite{banados}. For BTZ black holes, an analytical closed form
for the quasinormal frequencies was derived \cite{cardoso1}, and it was
verified that the quasinormal frequencies correspond exactly to the poles of
retarded correlation functions in the dual $(1+1)$-dimensional CFT
\cite{birm1}. It was also suggested in Ref. \cite{son1} that the relation
between quasinormal modes and singularities of correlation functions should
also hold for scalar fields in higher-dimensions, as far as the frequencies
are computed by imposing Dirichlet boundary conditions on such fields at AdS
infinity.

In the meantime, two fundamental difficulties arise when considering
gravitational and/or electromagnetic perturbations of AdS black holes,
particularly in higher dimensional spacetimes. The first problem is related
to the arbitrariness in the choice of gauge-invariant perturbation fields. In
fact, there is an infinity of gauge-invariant combinations of metric (or
vector potential) fluctuations that can be used as fundamental variables
governing the gravitational (or electromagnetic) perturbations. The second
problem is related to the ambiguity in defining appropriate boundary
conditions for the quasinormal modes. A traditional way to face such
arbitrariness is opting for master variables that lead to equations
generalizing those for perturbations in asymptotically flat spacetimes. That
is to say, variables are chosen in such a way to put the radial part of the
fundamental equations into a Schr\"odinger-like form. From now on, the
corresponding master variables shall be called the Regge-Wheeler-Zerilli
(RWZ) variables.\footnote{In the first study of gravitational QNM in AdS
spacetimes, Cardoso and Lemos \cite{cardoso2} used the same kind of variables
as the early works in asymptotically flat spacetimes by Regge and Wheeler
\cite{reggewheeler}, and by Zerilli\cite{zerilli}.}
With such a choice of variables, it was investigated gravitational and/or
electromagnetic perturbations of the Schwarzschild-AdS
\cite{cardoso2,cardoso3,konoplya4,cardoso4,lopez,musiri3,friess1},
Reissner-Nordstr\"om-AdS \cite{berti1,natario,shu}, and Kerr-AdS
\cite{giammatteo2} black holes, as well as the perturbations of black holes
with non-spherical topologies \cite{birm2,kout2}, including the
plane-symmetric ones \cite{cardoso5,miranda1,miranda2}. Analogously to the
massless scalar field case, in all of these works the quasinormal modes were
computed by imposing Dirichlet boundary conditions on the master fields at
infinity. Alternative boundary conditions for the same
Regge-Wheeler-Zerilli variables have been discussed in Refs.
\cite{moss,micha}.

A different route was taken by N\'u\~{n}ez and Starinets \cite{nunez}, who
defined the quasinormal frequencies of a perturbation in an asymptotically
AdS spacetime as ``the locations in the complex frequency plane of the poles
of the retarded correlator of the operators dual to that perturbation''. To
compute the real-time correlation functions, they suggested using the
Lorentzian AdS/CFT prescription of Refs. \cite{son1, herzog1}.
The quasinormal-mode definition supplied by N\'u\~{n}ez and Starinets was
explored in Ref. \cite{kovtun1}, where a new set of
fundamental variables was introduced to study electromagnetic and
gravitational perturbations of $(4+1)$-dimensional plane-symmetric black
holes
(or black branes, for short). It was shown there that the imposition of
Dirichlet boundary conditions on such a new set of gauge-invariant variables
at infinity leads exactly to the quasinormal frequencies
associated to the corresponding black branes. In the present work these
kind of fundamental variables shall be called the Kovtun-Starinets (KS)
variables.

An important consequence of the N\'u\~{n}ez-Starinets approach \cite{nunez}
is that the resulting quasinormal-mode spectra present a set of dispersion
relations, here called hydrodynamic QNM, that behave like diffusion, shear,
and sound wave modes in the long-wavelength, low-frequency limit
\cite{kovtun1}. These results are totally consistent with what is expected
from the CFT point of view, and they provide a non-trivial test of the
AdS/CFT correspondence. It is also worth noticing that neither the
electromagnetic diffusion mode nor the gravitational sound wave mode are
obtained by imposing Dirichlet boundary conditions on the RWZ
master variables. For Schwarzschild-AdS and topological-AdS
$(3+1)$-dimensional black holes, it was only possible to obtain sound wave
modes in the gravitational quasinormal spectra by requiring that a specific
combination of the master field and its derivative vanishes at infinity
\cite{micha,siopsis3,alsup}.

\subsection{The present work}

\subsubsection{General procedure}

In this work the definition of QNM given by N\'u\~nez and Starinets
\cite{nunez} is applied to compute the quasinormal frequencies associated to
electromagnetic and gravitational perturbations of $(3+1)$-dimensional
plane-symmetric AdS black holes. The overall procedure is similar to that of
Ref. \cite{kovtun1} and consists of the following steps:

\begin{itemize}

\item[(1)] Initially the translation invariance of the static plane-symmetric
AdS spacetimes is used to Fourier transform the fluctuation fields with
respect to time and to the two Cartesian coordinates $(x,y)$ of the plane.

\item[(2)] With the spatial wave vector chosen to be in the $y$-direction,
both the electromagnetic and the gravitational perturbation fields are
separated into two sets according to their behavior under the transformation
$x\rightarrow -x$: odd (axial, or transverse), and even (polar, or
longitudinal) perturbations. 

\item[(3)] Each sector of perturbation fields is governed by a set of
linearized differential equations. In all of the cases studied here, the
complete set of perturbation equations can be decoupled in order to obtain a
unique second-order differential equation, which is the fundamental equation
of that perturbation sector. The fundamental equations are written in
terms of gauge-invariant combinations of the perturbation fields, extending
the original definitions of Kovtun-Starinets variables \cite{kovtun1} to
$(3+1)$-dimensional spacetimes.

\item[(4)] Then, the standard AdS/CFT prescription of Ref. \cite{son1} is
applied to express the real-time $R$-symmetry current and stress-energy
tensor correlators in terms of quantities which represent the asymptotic
behavior of perturbations near the AdS-space boundary.  Such a procedure
shows that the imposition of Dirichlet boundary conditions on
Kovtun-Starinets variables at infinity leads to the poles of the CFT
correlation functions, and therefore, according to the N\'u\~{n}ez-Starinets
definition of QNM, to the quasinormal spectra of the plane-symmetric AdS
black holes.

\item[(5)] With well defined boundary conditions and a set of decoupled
fundamental equations, the hydrodynamical limit of the QNM spectra is then
analyzed. This limit is reached for perturbation modes in which the frequency
and the wavenumber are much smaller than the Hawking temperature of
the black hole.

\item[(6)] The last step is numerically compute the electromagnetic
and gravitational quasinormal dispersion relations for different
blackhole parameters. For such a purpose, the Horowitz-Hubeny
method \cite{horowitz1}, which reduces the problem of finding
QNM frequencies to that of obtaining the roots of infinite polynomial
equations, is used.

\end{itemize}

\subsubsection{Main results}

Among the new results found in the present work, it is worth
mentioning the following ones.
\begin{itemize}
\item First, the derivation of the electromagnetic diffusion mode and the
gravitational sound wave mode is  performed by means of a traditional QNM
calculation. These modes were earlier obtained by Herzog \cite{herzog2,
herzog3}, who utilized the AdS/CFT prescription \cite{son1} to directly
compute the hydrodynamic limit of the CFT $R$-symmetry current and
stress-energy tensor correlators. 

\item Second, it is found that the procedure of imposing Dirichlet boundary
conditions on the gauge-invariant KS variables breaks the
isospectrality between the axial and polar electromagnetic QNM, that follows
from RWZ variables. As a result, the polar electromagnetic
quasinormal modes are totally new, since the KS variable with
Dirichlet boundary condition yields a different spectrum when compared to the
RWZ variable with the same kind of boundary conditions. Regarding
to the electromagnetic axial perturbations, the dispersion relations found
here enlarge previous results of Ref. \cite{cardoso5}. 

\item  Third, it is found in addition that the complete spectra of
electromagnetic QNM present a tower of purely damped modes which tend to
Matsubara frequencies characteristic to bosonic systems in the
long-wavelength regime. However, these quasinormal modes do not exist for all
wavenumbers. In fact, there is a saturation value for the wavenumber above
which the electromagnetic purely damped modes disappear.

\item  Fourth, another result to be mentioned is the difference between the
spectrum of the gravitational polar perturbations, computed by using the
KS variable, and that obtained using the RWZ master
variable \cite{miranda1}. The differences are specially significant when the
fluctuation wavenumber is of the same order of the magnitude of the
blackhole temperature. 

\item And last but not least, the dispersion relations calculated here
complete the previous results for axial gravitational QNM of
$(3+1)$-dimensional plane-symmetric AdS black holes \cite{cardoso5,miranda1}.

\end{itemize}

\subsubsection{Structure of the paper}

The layout of the present article is as follows. Sect. \ref{m2branes}
contains a brief summary of the relation between the plane-symmetric
$\mbox{AdS}_{4}$ black holes and the eleven-dimensional supergravity solution
associated with a stack of $N$ M2-branes. In the sequence a detailed study of
the electromagnetic quasinormal modes is performed (Sect. \ref{qnm-eletro}):
The basic equations are obtained in Sect. \ref{flut-eletro} and the
connection between the blackhole perturbations and the CFT $R$-symmetry
currents is explored in Sect. \ref{green-electro}; the hydrodynamic modes of
the electromagnetic perturbations are studied in Sect.
\ref{hydro-eletro}, and the general dispersion relations of
electromagnetic QNM are reported in Sect. \ref{dispersion-eletro}. The
gravitational quasinormal modes are studied in Sect. \ref{grav-qnm}:
The basic equations are obtained in Sect. \ref{flut-grav}; 
\ref{green-gravit} is devoted to investigate the relation between the
gravitational QNM and the stress-energy tensor correlators in the
holographic CFT; the hydrodynamic modes of the gravitational
perturbations are studied in Sect. \ref{hydro-grav}, and the numerical
results for the dispersion relations of the remaining gravitational QNM
are presented in Sect. \ref{dispersion-gravit}. The article is completed, in
Sect. \ref{consid-final}, with the analysis and interpretation of the main
results.

\subsection{Notation and conventions}

Natural units are going to be used throughout this paper, i.e., the speed of
light $c$, Boltzmann constant $k_B$, and Planck constant $\hbar$ are all set
to unity, $c=k_{B}=\hbar=1$. Regarding to notation, capital Latin indices
$M,\,N,\,...$ vary over the coordinates of the whole AdS spacetime, while
Greek indices $\mu,\,\nu,\,...$ label different coordinates at the boundary,
and small Latin indices $i,\,j,\,...$ vary only over the spacelike
coordinates at the boundary. The convention for the metric signature and for
all the definitions of curvature tensors follow Ref. \cite{misner}.

\section{M2-branes and the plane-symmetric black holes}
\label{m2branes}

\subsection{The background spacetime}

Since the QNM definition of N\'u\~{n}ez-Starinets \cite{nunez}, that is
adopted in this work, makes heavy use of the relation between AdS black holes
and conformal field theories at finite temperature, it becomes important to
review here how the plane-symmetric AdS$_4$ black holes arise in the context
of the AdS/CFT conjecture.\footnote{The brief summary presented in this
section is based on material found in Refs. \cite{herzog2, aharo, herzog4}.}

A fundamental role in the AdS/CFT correspondence\footnote{The interest here
is the AdS$_4$/CFT$_3$ correspondence.} is played by
extended two-dimensional objects known as M2-branes \cite{townsend}.
The world-volume theory
of $N$ M2-branes is a $(2+1)$-dimensional non-Abelian Yang-Mills theory
which presents $\mathcal{N}=8$ supersymmetries in addition to a
$SU(N)$ gauge group. The coupling constant of the theory
flows to strong coupling in the infrared limit, and it is believed that
the flow is to an infrared-stable fixed point that describes
a superconformal field theory \cite{seiberg}.
This CFT also has an emerging $R$-charge symmetry which
is expanded to $SO(8)$.

From the supergravity point of view, a stack of $N$ M2-branes
is described by a nonextremal solution to the
supergravity equations of motion,
characterized by the metric \cite{herzog2, horowitz2, itzhaki}
\begin{equation}
ds^2=H^{-2/3}(\widetilde{r})\left[-\mathfrak{h}(\widetilde{r})dt^2+dx^{2}+
dy^{2}\right]+H^{1/3}(\widetilde{r})\left[\mathfrak{h}^{-1}(\widetilde{r})
d\widetilde{r\,}^{2}+\widetilde{r\,}^2 d\Omega_{7}^{2}\right],
\label{metM2brana}
\end{equation}
where
\begin{equation}
H(\widetilde{r})=1+\left(\frac{R}{\widetilde{r}}\right)^{6}
\qquad\mbox{and}\qquad\mathfrak{h}(\widetilde{r})=1-
\left(\frac{\widetilde{r}_{0}}{\widetilde{r}}\right)^{6},
\label{hdef1}
\end{equation}
and by a four-form field whose dual Hodge is given by
\begin{equation}
\star F_{4}=F_{7}=6R^{6}
\mbox{Vol}(S^{7}){\boldsymbol{\varepsilon}},
\end{equation}
where $\boldsymbol{\varepsilon}$ stands for the Levi-Civita
tensor on $S^{7}$.
According to the AdS/CFT correspondence \cite{maldacena1,gubser1,witten1},
the $(2+1)$-dimensional $\mathcal{N}=8$ CFT is dual to M-theory on the
background spacetime \eqref{metM2brana}. Furthermore, the quantization
condition on the ${F}_{4}$ flux connects the parameter $R$ to the number of
branes $N$ \cite{klebanov1}:
\begin{equation}
R^{9}\pi^{5}=N^{3/2}\kappa_{11}^{2}\sqrt{2},
\label{relparam}
\end{equation}
where $\kappa_{11}$ is the gravitational coupling strength in
$(10+1)$-dimensional supergravity.

In the large $N$ limit ($N\gg 1$), one can consider only the near-horizon
region ($\widetilde{r}\ll R$) of the spacetime \eqref{metM2brana}.
Function $H(\widetilde{r})$ then reduces to
$H(\widetilde{r})=R^{6}/\widetilde{r\,}^{6}$. Moreover,
defining a new radial coordinate by $r=\widetilde{r\,}^{2}/2R$,
metric \eqref{metM2brana} becomes
\begin{equation}
ds^2=\frac{4r^2}{R^2}\left[-\mathfrak{h}(r)dt^2+
dx^{2}+dy^{2}\right]+\frac{R^2}{4r^2}
\frac{dr^2}{\mathfrak{h}(r)}
+R^{2}d\Omega_{7}^{2}.
\label{nearhorizon}
\end{equation}
 The AdS part of the metric \eqref{nearhorizon},
associated to the coordinates $\{t,x,y,r\}$, is identical to the
solution of Einstein equations with negative cosmological term
corresponding to a $(3+1)$-dimensional plane-symmetric AdS black hole
\cite{lemos0,lemos1, huang, cai}: 
\begin{equation}
ds^{2}=-f(r)dt^{2}+f(r)^{-1}dr^{2}+
\frac{r^{2}}{L^{2}}(dx^{2}+dy^{2}),
\label{background}
\end{equation}
where the horizon function $f(r)$ is given by
\begin{equation}
f(r)=\left(\frac{r}{L}\right)^{2}\,\mathfrak{h}(r)=
\left(\frac{r}{L}\right)^{2}\left(1-\frac{r_{0}^{3}}{r^{3}}\right),
\label{fdef1}
\end{equation}
and the seven-sphere radius $R$ has been rewritten as $R=2L$, with $L$ now
representing the AdS radius of the spacetime \eqref{background}. Parameters
$r_0$ and $L$ are related to the blackhole Hawking temperature $T$ by
\begin{equation}
T=\frac{3}{4\pi}\frac{r_{0}}{L^{2}}.
\label{hawk-temperature}
\end{equation}

\subsection{Normalization of the field action}

The full theory is the eleven-dimensional supergravity on
$\mbox{AdS}_{4}\times S^{7}$, and the existence of a compact seven-sphere
enables one to consistently reduce the theory to Einstein-Maxwell theory on
$\mbox{AdS}_{4}$ \cite{herzog4, duff1, berenstein1}. The main objective in
summarizing such a procedure here is to make explicit the dependence of the
action for the fields in the AdS$_{4}$ spacetime on the number of colours
$N$, which is one of the  parameters characterizing the holographic CFT. 

Upon Kaluza-Klein dimensional reduction, the Maxwell gauge field
$A_{\ss{M}}$ arises from a combination of metric and ${F}_{4}$ form
perturbations in the eleven-dimensional supergravity. This field corresponds
to a $U(1)$ subgroup of the $SO(8)$ symmetry group of the complete spacetime
\eqref{metM2brana}. The mechanism of dimensional reduction also furnishes
the $(3+1)$-dimensional Einstein-Maxwell action with a negative cosmological
constant $\Lambda=-3/L^{2}$:
\begin{equation}
S=\frac{1}{2\kappa_{4}^{2}}\int d^{4}x\sqrt{-g}\left(\mathcal{R}
+\frac{6}{L^{2}}-L^{2}F_{\ss{MN}}F^{\ss{MN}}\right),
\label{acaocompleta}
\end{equation}
where $\mathcal{R}$ denotes the Ricci scalar and $F_{\ss{MN}}$ is the
electromagnetic strength tensor, and for the purposes of the present
analysis the electromagnetic Lagrangian $\mathcal{L}_{em} \sim
F_{\ss{MN}}F^{\ss{MN}}$ is considered as a perturbation on the gravitational
Lagrangian $\mathcal{L}_{gr} \sim \mathcal{R} +{6}/{L^{2}} $. It is
assumed that the gravitational coupling constants in four and eleven
dimensions are related by means of the seven-sphere volume \cite{herzog4},
\begin{equation}
\frac{1}{2\kappa_{4}^{2}}=\frac{R^{7}
\mbox{Vol}(S^{7})}{2\kappa_{11}^2}.
\end{equation}
Then, considering that the volume of a unitary seven-sphere
is $\mbox{Vol}(S^{7})=\pi^{4}/3$, and using the standard
normalization \eqref{relparam} for $\kappa_{11}$,
it is found 
\begin{equation}
\frac{1}{2\kappa_{4}^{2}}=\frac{\sqrt{2}N^{3/2}}{24\pi L^{2}}.
\label{gravity-const}
\end{equation}
Action \eqref{acaocompleta} with the gravitational constant $\kappa_{4}$
given in terms of the number of colours $N$ and of the anti-de Sitter
radius $L$ is the desired result, which is needed for the development of
the present work.

\section{Electromagnetic quasinormal modes}
\label{qnm-eletro}

\subsection{Perturbation equations}
\label{flut-eletro}

In the AdS/CFT context, the electromagnetic field in the AdS bulk couples to
the CFT $R$-symmetry currents at the spacetime boundary. Hence, in order to
construct the current-current two-point correlation functions in the CFT, it
is necessary to consider fluctuations of the gauge field $A_{\ss{M}}$. Such a
field is implicitly defined  by 
\begin{equation}
F_{\ss{MN}}=\partial_{\ss{M}}A_{\ss{N}}-
\partial_{\ss{N}}A_{\ss{M}},
\label{strength}
\end{equation}
with $F_{\ss{MN}}$ satisfying equations of motion derived from the action
\eqref{acaocompleta}. Therefore, considering the electromagnetic field as a
perturbation on the background spacetime of metric \eqref{background}, the
resulting equations of motion for $A_{M}$ are the usual Maxwell equations
\begin{equation} 
\partial_{\ss{M}}\left(\sqrt{-g}g^{\ss{MA}}
g^{\ss{NB}}F_{\ss{AB}}\right)=0,
\label{maxwell}
\end{equation}
where $g_{\ss{MN}}$ stands for the metric components given by
\eqref{background}.

When looking for solutions to Eqs. \eqref{maxwell}, by taking into account
the isometries of the background metric \eqref{background}, it is convenient
to decompose the gauge field in terms of Fourier transforms as follows
\begin{equation}
A_{\ss{M}}(t,x,y,r)=\frac{1}{(2\pi)^{3}}\int{\! d\omega\, dk_{x}\, dk_{y}\,}
e^{-i\omega t+ik_{x}x+ik_{y}y}\widetilde{A}_{\ss{M}}(\omega,k_{x},k_{y},r). 
\label{EMfourier}
\end{equation}
Furthermore, without loss of generality, in the plane-symmetric background
spacetime \eqref{background} one may choose the wave three-vector $k$ in
the form $k_\mu=(k_0,k_x,k_y)=(-\omega,0,q)$. This is carried out
through an appropriate rotation in the $x-y$ plane, in such a way that the
Fourier modes of the gauge field propagate along the $y$ direction only. With
such a choice, the electromagnetic perturbations $ A_{\ss{M}}$ can be split
into two independent sets according to their behavior under parity operation,
$x\rightarrow -x$:
\begin{itemize}
\item Axial (odd, or transverse) perturbations:
$A_{x}$;
\item Polar (even, or longitudinal) perturbations:
$A_{t}$, $A_{y}$, $A_{r}$.
\end{itemize}
Since these two sets of perturbations are orthogonal sets, they can
be studied separately, as it is done in the following.

\subsubsection{Equations for axial perturbations}

Axial electromagnetic perturbations are governed by the transverse component
of Maxwell equations \eqref{maxwell}, which gives
\begin{equation}
f\frac{d^2 A_{x}}{dr^2}+\frac{df}{dr}\frac{d A_{x} }{dr}+
\left(\frac{\omega^2 r^2-q^2 L^{2}f}{fr^2}\right) A_{x}=0,
\label{axeletro1}
\end{equation}
where, to simplify notation, the tilde was dropped, $\widetilde A_x
\rightarrow A_x$. Moreover, it follows from Eqs. \eqref{strength} and
\eqref{EMfourier}, together with $k_{\mu}=(-\omega,0,q)$, that  $A_{x}$ is
proportional to the transverse component of the electric field: $
E_{x}=i\omega A_{x}$. Therefore, being a gauge-invariant quantity, $A_{x}$ is
also a good candidate as master variable for axial perturbations. In fact, it
is possible to cast Eq. \eqref{axeletro1} into a Schr\"odinger-like form
\cite{cardoso5}
\begin{equation}
\left(\frac{d^{\,2}}{dr_{\ast}^{2}}+\omega^{2}\right)
\Psi^{\ss{(-)}}=f\left(\frac{qL}{r}\right)^{2}\,
\Psi^{\ss{(-)}},
\label{ondaeletro1}
\end{equation}
where $\Psi^{\ss{(-)}}(r)=A_{x}(r)$, and the tortoise coordinate $r_{\ast}$
is defined in terms of the radial coordinate $r$ by
\begin{equation}
\frac{dr}{dr_{\ast}} = {f(r)} \label{tortoise}.
\end{equation}

For the present purposes it is convenient to change coordinates to the
inverse radius $u=r_{0}/r$, and then, by writing Eq. \eqref{ondaeletro1} in
terms of $E_{x}$ it results
\begin{equation}
E_{x}^{''}+\frac{\mathfrak{h}^{'}}{\mathfrak{h}}E_{x}^{'}+
\frac{\mathfrak{w}^2-\mathfrak{q}^2\mathfrak{h}}{\mathfrak{h}^2}E_{x}=0, 
\label{fieletro}
\end{equation}
where the primes denote derivation with respect to the variable
$u$, and $\mathfrak{w}$ and $\mathfrak{q}$ are respectively the
normalized frequency and wavenumber, defined by
\begin{equation}
\mathfrak{w}=\frac{3\omega}{4\pi T}\quad\qquad\mbox{and}
\qquad\quad\mathfrak{q}=\frac{3q}{4\pi T},
\label{wq-normalizados}
\end{equation}
with $T$ being the Hawking temperature of the black hole, given by Eq.
\eqref{hawk-temperature}. Function $\mathfrak{h}$ is obtained from Eqs.
\eqref{hdef1}, or from Eq. \eqref{fdef1}, and in terms of the variable
$u=r_0/r$ reads
\begin{equation}
 \mathfrak{h}\equiv \mathfrak{h}(u)=1-u^3\,.
\label{hdef2}
\end{equation} 
Quantity $E_x$ shall be the fundamental gauge-invariant
variable to be used in the present analysis of the QNM modes for axial
electromagnetic perturbations.

\subsubsection{Equations for polar perturbations}

Differently from the axial electromagnetic perturbations, the components of
the gauge field $A_{\ss{M}}$ corresponding to the set of polar perturbations
are not gauge invariant. The gauge freedom can then be used in order to
simplify the relevant equations of motion. In fact, the invariance of Maxwell
equations under the gauge transformation $A_{\ss{M}}\rightarrow
A_{\ss{M}}+\partial_{\ss{M}}\lambda$ allows choosing $\lambda$ in such a way
that one of the components $A_t,\,A_y,$ or $A_r$ vanishes. For instance, it
is possible to work in the so-called radial gauge, in which $A_r=0$
\cite{herzog2}. In this gauge, Maxwell equations \eqref{maxwell}
corresponding to the polar perturbations are 
\begin{equation}
\omega r^{2}\frac{d}{dr}A_{t}+qL^{2}f\frac{d}{dr}A_{y}=0,
\label{eqeletro1}
\end{equation}
\begin{equation}
r^{2}\frac{d^2}{dr^2}A_{t}+2r\frac{d}{dr}A_{t}-\frac{L^{2}}{f}
\left(q\omega A_{y}+q^{2}A_{t}\right)=0,
\label{eqeletro2}
\end{equation}
\begin{equation}
f\frac{d^2}{dr^2}A_{y}+\frac{df}{dr}\frac{d}{dr}A_{y}
+\frac{1}{f}\left(\omega qA_{t}+\omega^{2}A_{y}\right)=0.
\label{eqeletro3}
\end{equation}
Notice that this set of equations does not constitute a linearly independent
system, and any subset composed by two of such equations determines the two
remaining unknown components of the gauge field, $A_t$ and $A_y$. 

From now on, $A_t$ or $A_y$ could be adopted as a primary variable 
and equations \eqref{eqeletro1}-\eqref{eqeletro3} could be
decoupled in order to find a unique differential equation for one of
these functions. However, to avoid the inconvenient of dealing with
gauge dependent quantities, it is interesting to use
the electric field components, which are gauge-invariant quantities.
Even though this choice eliminates gauge-dependent potential fields,
a residual ambiguity is left: from the electric field components
$E_{r}=dA_{t}/dr$ and $E_{y}=i(q A_{t}+\omega A_{y})$, what is the best choice?

The answer to the last question is not simple and both of the possible
answers have been tried in the literature. For instance, inspired by
preceding works studying blackhole perturbations in asymptotically flat
spacetimes \cite{ruffini}, Cardoso and Lemos opted for the radial component
$E_r$ in studying QNM of Schwarzschild-AdS black holes \cite{cardoso2},
and plane-symmetric AdS black holes \cite{cardoso5}. Introducing a new
variable
$\Psi^{\ss{(+)}}(r)=r^{2}E_{r}(r)$ and using Eqs. \eqref{eqeletro1} and
\eqref{eqeletro2}, they were able to reduce the system of equations into a
unique ordinary differential equation of Schr\"odinger type
\begin{equation}
\left(\frac{d^{\,2}}{dr_{\ast}^{2}}+\omega^{2}\right)
\Psi^{\ss{(+)}}=f\left(\frac{qL}{r}\right)^{2}
\Psi^{\ss{(+)}},
\label{ondaeletro2}
\end{equation}
which has the same form as the fundamental equation for axial perturbations,
Eq. \eqref{ondaeletro1}. An interesting consequence of this fact is that the
QNM spectra for both the axial and polar perturbations, with the same
boundary conditions, are identical \cite{cardoso2,cardoso5}. As a matter of
fact, an open question left behind in the early works computing
electromagnetic QNM of AdS black holes is the lack of a convincing physical
justification for the choice of Dirichlet boundary conditions at infinity for
both the polar and the axial perturbations. Such an issue will be considered
in the sequence of this work.

 As far as one is interested in computing the polar electromagnetic
quasinormal frequencies, it will be shown in Sect. \ref{green-electro} that
$E_y$ is  more appropriate as a fundamental variable than $E_r$. This was
the choice made, for instance, by Kovtun and Starinets in studying QNM of
black branes in $(4+1)$-dimensional spacetimes \cite{kovtun1}. Following
these
authors, in the present work $E_y$ is adopted as the fundamental variable
to be used to determine the QNM spectrum of polar electromagnetic
perturbations. Having made this choice, one then writes equations in terms of
the independent variable $u=r_0/r$. With this, Eqs. \eqref{eqeletro1} and
\eqref{eqeletro2} written for $E_{y}$ lead to
\begin{equation}
E_{y}^{''}+\frac{\mathfrak{w}^{2}\mathfrak{h}^{'}}{\mathfrak{h}\left(
\mathfrak{w}^2-\mathfrak{q}^{2}\mathfrak{h}\right)}E_{y}^{'}+
\frac{\left(\mathfrak{w}^2-\mathfrak{q}^{2}\mathfrak{h}
\right)}{\mathfrak{h}^2}E_{y}=0,
\label{yeletro}
\end{equation}
where $\mathfrak{w}$ and $\mathfrak{q}$ are respectively the normalized
frequency and wavenumber, defined by Eqs. \eqref{wq-normalizados},
and $\mathfrak{h}$ is given by Eq. \eqref{hdef2}.

According to the N\'u\~{n}ez-Starinets QNM definition \cite{nunez}, once one
has found the fundamental perturbation equations for a field on the AdS
spacetime, the next step is establishing explicit relations between the
perturbation variables and the corresponding retarded Green functions in the
holographic CFT. It is exactly from these relations that will emerge the
boundary conditions to be imposed on the perturbation fields at infinity,
viz, the conditions that lead to the singularities of the two-point
correlation functions in the boundary field theory, and consequently to the
quasinormal frequencies of the fluctuation modes. Such a task is performed in
what follows.

\subsection{$R$-current correlation functions}
\label{green-electro}

In the case of electromagnetic perturbations, the AdS/CFT correspondence
\cite{maldacena1,gubser1,witten1} tells that, in the strong coupling, large
$N$ limit, the information on the thermal correlation functions of the
$R$-symmetry currents are encoded into the electric field components $E_{j}$
($j=x,y$), which are solutions to the differential equations \eqref{fieletro}
and \eqref{yeletro}, respectively. It can be shown that, close to the horizon
($u\approx 1$), such functions are given approximately by
$E_{j}=\mathfrak{h}^{\pm i\mathfrak{w}/3}$, where the negative (positive)
exponent corresponds to ingoing (outgoing) waves. Moreover, depending on the
sign of the exponent, the boundary values of the perturbation functions act
as sources of retarded or advanced Green functions in the dual CFT. To
compute the retarded two-point functions, one has to opt for the negative
exponent. It is also necessary to know the asymptotic form of the
perturbation functions close to the infinite boundary ($u\approx 0$). A
simple analysis shows that the solutions of equations \eqref{fieletro} and
\eqref{yeletro} which satisfy incoming-wave condition at horizon
present the
following behavior around $u=0$:
\begin{gather}
E_{x}=\mathcal{A}_{(x)}(\mathfrak{w},\mathfrak{q})
+...\;+\mathcal{B}_{(x)}(\mathfrak{w},\mathfrak{q})u+ ... \,,
\label{assintelet1}\\
E_{y}=\mathcal{A}_{(y)}(\mathfrak{w},\mathfrak{q})+...\,+
\mathcal{B}_{(y)}(\mathfrak{w},\mathfrak{q})u+...\, ,
\label{assintelet2}
\end{gather}
where ellipses denote higher powers of $u$ for each one of the independent 
solutions. Symbols $\mathcal{A}_{(j)}(\mathfrak{w},
\mathfrak{q})$ and $\mathcal{B}_{(j)}(\mathfrak{w},
\mathfrak{q})$, introduced in the above equations, stand for the connection
coefficients associated to the corresponding differential equations for
$E_x$ and $E_y$.

To proceed further and calculate the correlation functions, the
electromagnetic action at the boundary needs to be determined.
It is usual to split the action as $S_{\ss{EM}} =
S_{\ss{horizon}}+S_{\ss{boundary}}$. Using the equations of motion and the
preceding definitions it follows 
\begin{equation}
S_{\ss{boundary}}=\frac{\chi}{2}\;\underset{u\rightarrow 0}{\mbox{lim}}
\,\int\frac{d\mathfrak{w} d\mathfrak{q}}{(2\pi)^2}\left[\frac{\mathfrak{h}}{
\mathfrak{w}^{2}-\mathfrak{q}^{2}\mathfrak{h}}E_{y}^{'}(u,k)E_{y}(u,
-k)+\frac{\mathfrak{h}}{\mathfrak{w}^{2}}
E_{x}^{'}(u,k)E_{x}(u,-k)\right],
\label{acaoeletro}
\end{equation}
where
\begin{equation}
\chi=\frac{8\pi TL^{2}}{3\kappa_{4}^{2}}
=\frac{(2N)^{3/2}T}{9}
\end{equation}
is the electric susceptibility of the dual system \cite{herzog2,herzog4}.

In order to apply the Lorentzian AdS/CFT prescription of Ref. \cite{son1},
the asymptotic solutions \eqref{assintelet1} and \eqref{assintelet2} are used
to write the derivatives of the electric field in terms of the boundary
values of the three-vector potential $A_{\mu}^{0}(k)=A_{\mu}(u\rightarrow 0,
k)$. The $R$-current correlation functions $C_{\mu\nu}$ are proportional
to the coefficients of the terms containing the product
$A_{\mu}^{0}(k)A_{\nu}^{0}(-k)$ that appears in the action
\eqref{acaoeletro}. It is then found:
\begin{equation}
\begin{aligned}
&C_{tt}=\chi\frac{\mathfrak{q}^{2}}{\left(\mathfrak{w}^{2}
-\mathfrak{q}^{2}\right)}\frac{\mathcal{B}_{(y)}(\mathfrak{w},\mathfrak{q})}
{\mathcal{A}_{(y)}(\mathfrak{w},\mathfrak{q})},
&\qquad& C_{yy}=\chi\frac{\mathfrak{w}^{2}}{\left(\mathfrak{w}^{2}
-\mathfrak{q}^{2}\right)}\frac{\mathcal{B}_{(y)}(\mathfrak{w},\mathfrak{q})}
{\mathcal{A}_{(y)}(\mathfrak{w},\mathfrak{q})},\\
&C_{ty}=-\chi\frac{\mathfrak{w}\mathfrak{q}}
{\left(\mathfrak{w}^{2}-\mathfrak {q}^{2}\right)}\frac{\mathcal{B}_{(y)}
(\mathfrak{w},\mathfrak{q})}{\mathcal{A}_{(y)}(\mathfrak{w},\mathfrak{q})},
&\qquad& C_{xx}= \chi\frac{\mathcal{B}_{(x)}(\mathfrak{w},\mathfrak{q})}
{\mathcal{A}_{(x)}(\mathfrak{w},\mathfrak{q})}.\label{celetro}
\end{aligned}
\end{equation}
Moreover, for a $(2+1)$-dimensional CFT at finite temperature, the
current-current correlation functions can be written in terms of the
transverse and longitudinal self-energies
$\Pi^{T}(\mathfrak{w},\mathfrak{q})$ and
$\Pi^{L}(\mathfrak{w},\mathfrak{q})$, respectively (See Appendix
\ref{apen-correlations} for a summary of such relations). Hence, comparing
Eqs. \eqref{celetro} to Eqs. \eqref{tself-en}--\eqref{lself-en3} of Appendix
\ref{apen-correlations}, one finds
\begin{equation}
\Pi^{T}(\mathfrak{w},\mathfrak{q})=\chi\frac{\mathcal{B}_{(x)}(\mathfrak{w},
\mathfrak{q})}{\mathcal{A}_{(x)}(\mathfrak{w},\mathfrak{q})},
\qquad\qquad\Pi^{L}(\mathfrak{w},\mathfrak{q})=\chi
\frac{\mathcal{B}_{(y)}(\mathfrak{w},\mathfrak{q})}{
\mathcal{A}_{(y)}\mathfrak{w},\mathfrak{q})}.
\end{equation}
These results show that the retarded two-point correlation functions are
fully determined by the ratio between the connection coefficients of
equations \eqref{fieletro} and \eqref{yeletro}. Furthermore, the poles of
the thermal correlation functions are given by the zeros of the coefficients
$\mathcal{A}_{(x)}(\mathfrak{w},\mathfrak{q})$ and
$\mathcal{A}_{(y)}(\mathfrak{w}, \mathfrak{q})$. According to Ref.
\cite{nunez}, the poles of $C_{\mu\nu}$ define the electromagnetic QNM
frequencies of the black hole localized in the AdS spacetime. Such
frequencies are then obtained by imposing Dirichlet boundary conditions
on the electric field components $E_{x}$ and $E_{y}$ at $u=0$, with $E_{x}$
and $E_{y}$ being functions that satisfy also an incoming-wave condition at
the horizon.

\subsection{QNM and the gauge-invariant variables}

At this stage one could ask whether imposing Dirichlet conditions at the
boundary ($u=0$) on the Regge-Wheeler-Zerilli (RWZ) variables
$\Psi^{\ss{(\pm)}}$ would produce the same QNM spectra as the spectra
obtained by imposing the same boundary conditions onto the Kovtun-Starinets
(KS) variables $E_{x,y}$. For the transverse electromagnetic sector, the
answer to this question is quite easy to find. In fact, variables
$\Psi^{\ss{(-)}}$ and $E_{x}$ are proportional to each other, so that both of
the obtained QNM spectra, either using the KS variable or using the RWZ
variable,  are identical. In the case of polar electromagnetic perturbations,
equations for the KS variable $E_{y}$ and for the RWZ variable
$\Psi^{\ss{(+)}}$  do not have the same form and, in addition, $E_{y}$ and
$\Psi^{\ss{(+)}}$ are independent variables, so that the answer is not
immediate. In fact, as it is shown below (see Sects. \ref{hydro-eletro} and
\ref{dispersion-eletro}), the QNM spectrum obtained from  $E_{y}$ is
different from the QNM spectrum obtained from $\Psi^{\ss{(+)}}$.

\subsection{Dispersion relations for the hydrodynamic QNM}
\label{hydro-eletro}

The hydrodynamic limit of perturbations corresponds to the small frequency
($\mathfrak{w}\ll 1$) and small wavenumber ($\mathfrak{q}\ll 1$) region of
the spectrum of the respective Fourier modes. In general, the quasinormal
modes can be classified according the behavior of the dispersion relations in
the hydrodynamic limit, and in this respect there are two classes. There is a
set of QNM for which the frequency $\mathfrak{w}(\mathfrak{q})$ vanishes when
$\mathfrak{q}\rightarrow 0$. Such modes are named here hydrodynamic
quasinormal modes. But there is another kind of QNM for which the
corresponding frequency in the long-wavelength limit is nonzero. To
distinguish these two kind of modes from each other, the modes belonging to
the later kind are denominated non-hydrodynamic quasinormal modes. In this
section, the dispersion relations of the electromagnetic hydrodynamic QNM are
studied by means of analytical and numerical methods. The electromagnetic
non-hydrodynamic QNM shall be object of study in the next section.

From the CFT point of view, it is expected that at least one of the
electromagnetic QNM should show the typical behavior of a diffusion mode
in the hydrodynamic limit. Trying to find such a mode, one then
looks for solutions to Eqs. \eqref{fieletro} and \eqref{yeletro} in
the form of power series in $\mathfrak{w}$ and $\mathfrak{q}$,
under the assumption $\mathfrak{w}\sim\mathfrak{q}$. Written
in terms of the variables $F_{j}=\mathfrak{h}^{i\mathfrak{w}/3}E_{j}$
(for $j=x,y$), which are more appropriate for the present analysis,
perturbation equations \eqref{fieletro} and \eqref{yeletro} may be cast as
\begin{equation}
F_{j}^{''}+\frac{u^{2}}{\mathfrak{h}}\left(2i\mathfrak{w}-3a_{j}
\right)F_{j}^{'}+\frac{1}{\mathfrak{h}^{2}}\left[i\mathfrak{w}(2u+u^{4}
-3a_{j}u^{4})+\mathfrak{w}^{2}(1-u^{4})-\mathfrak{q}^{2}\mathfrak{h}
\right]F_{j}=0,
\label{perturbativa1}
\end{equation}
where $a_{x}=1$ and $a_{y}=\mathfrak{w}^{2}/(\mathfrak{w}^{2}-
\mathfrak{q}^{2}\mathfrak{h})$.
After relabelling parameters as 
$\mathfrak{w}\rightarrow\lambda\mathfrak{w}$ and  $\mathfrak{q}
\rightarrow\lambda\mathfrak{q}$ with $\lambda\ll 1$,
it is assumed that solutions of Eqs. \eqref{perturbativa1}
can be expanded in the form
\begin{equation}
F_{j}(u)=F_{j}^{0}(u)+\lambda F_{j}^{1}(u)+\lambda^{2}
F_{j}^{2}(u)+...,\;
\label{expansaoeletro}
\end{equation}
where the coefficients $F_{j}^{\alpha}(u)$, with $\alpha = 0,1,2, ...,$
represent arbitrary functions of variable $u$, and which are also homogeneous
functions of degree $\alpha$ on $\mathfrak{w}$ and $\mathfrak{q}$.
 
The boundary condition of being ingoing waves at the horizon imposed on
$E_j$, when translated to the new functions $F_j$, implies their dominant
terms in expansion \eqref{expansaoeletro} must assume constant values close
the horizon ($u\approx 1$). Then, in terms of the expansion
\eqref{expansaoeletro} one has the following conditions
\begin{equation}
F_{j}^{0}(1)=\mbox{constant},\qquad F_{j}^{1}(1)=F_{j}^{2}(1)=...=0.
\label{contorno1}
\end{equation}
It is now possible to solve Eqs. \eqref{perturbativa1} order by order and,
after imposing the boundary conditions given in Eqs. \eqref{contorno1}, the
following expansions are found:
\begin{equation}
E_{x}=C_{x}\mathfrak{h}^{-i\mathfrak{w}/3}\bigg[1-
i\mathfrak{w}\frac{\sqrt{3}}{3}\left(\frac{\pi}{3}
-\arctan{\frac{1+2u}{\sqrt{3}}}\right)+\frac{i\mathfrak{w}}{2}
\ln{\frac{1+u+u^2}{3}}+\mathcal{O}(\mathfrak{w}^{2})\bigg],
\label{fiserie}
\end{equation}
\begin{equation}
E_{y}=C_{y}\mathfrak{h}^{-i\mathfrak{w}/3}\bigg[1+
\frac{i\mathfrak{q}^{2}}{\mathfrak{w}}(1-u)-i\mathfrak{w}
\frac{\sqrt{3}}{3}\left(\frac{\pi}{3}-\arctan{\frac{1+2u}{
\sqrt{3}}}\right)+\frac{i\mathfrak{w}}{2}\ln{\frac{1+u+u^2}{3}}
+\mathcal{O}(\mathfrak{w}^{2})\bigg],\label{yserie}
\end{equation}\\
where $C_{x}$ and $C_{y}$ are arbitrary normalization constants.
One finds from Eq. \eqref{fiserie} no solution satisfying the Dirichlet
condition at the AdS spacetime boundary, $E_x(0) =0$, and which is at the
same time compatible with the hydrodynamic approximation
$\mathfrak{w},\mathfrak{q}\ll 1$. This means there is no axial
electromagnetic hydrodynamic QNM, and no $R$-charge diffusion in the
transverse direction to the spatial wave vector, as expected from the CFT
point of view. On the other hand, the condition $E_{y}(0)=0$ and Eq.
\eqref{yserie} lead to\footnote{This result was also found through direct
calculation of the hydrodynamic limit of correlation functions by Herzog
\cite{herzog2,herzog3}. The comparison to that result is in fact a test for
the analysis performed in Sect. \ref{green-electro}.}
\begin{equation}
\mathfrak{w}=-i\mathfrak{q}^{2}
\qquad\Longrightarrow\qquad
\omega=-\frac{3i}{4\pi T}q^2,
\end{equation}
from where one can read the diffusion coefficient $D=3/4\pi T$. 
It is worth noticing that this diffusion mode is not found if one uses the
RWZ master variable $\Psi^{\ss{(+)}}$ instead of $E_{y}$.
\FIGURE{
%\begin{figure}[t]
\centering\epsfig{file=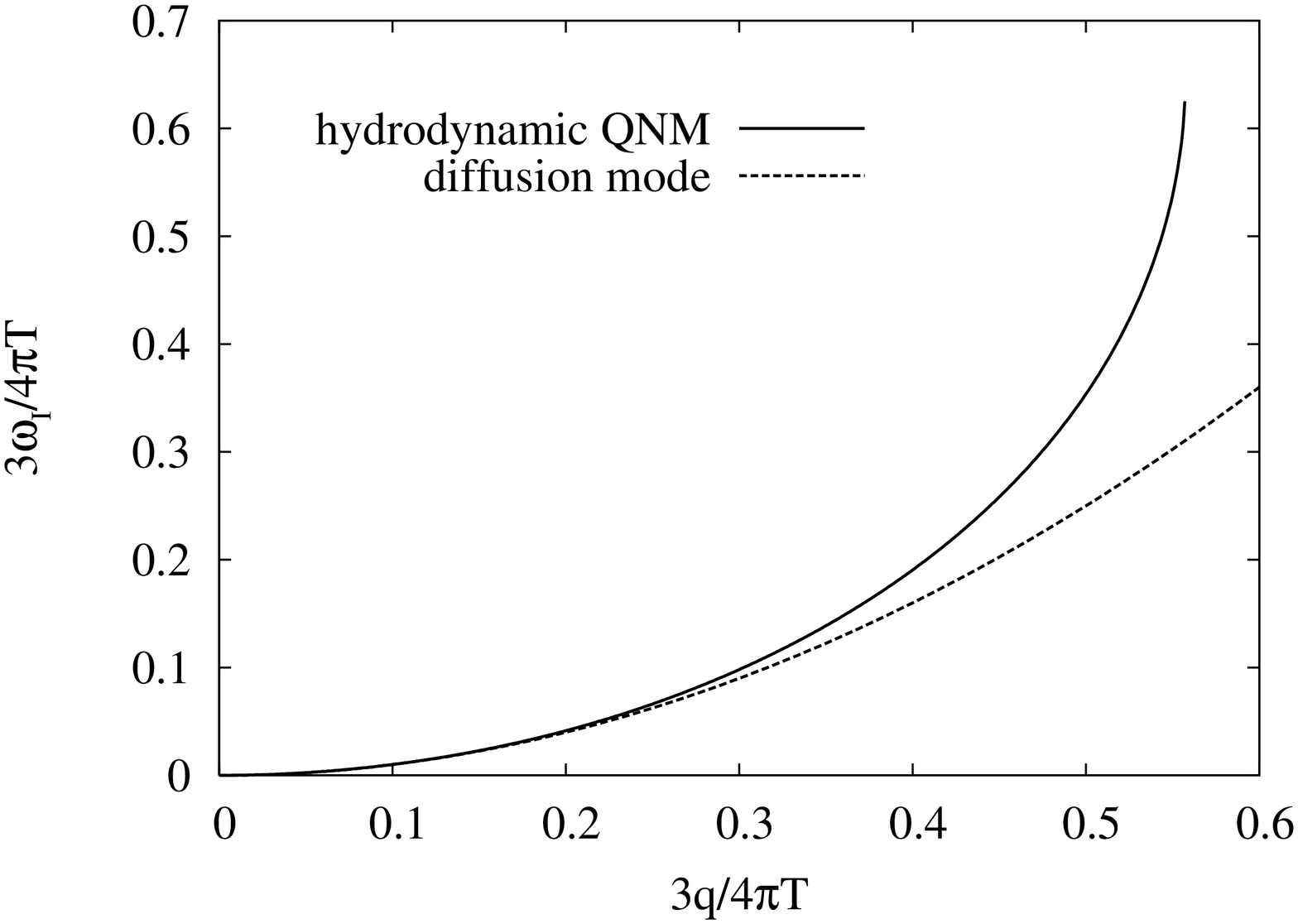, height=9.41cm,
width=6.0cm, angle=270}
\caption{The dispersion relation for the only electromagnetic hydrodynamic
QNM (solid line), which is purely damped, $\mathfrak{w}=-i\mathfrak{w}_{I}$,
and corresponds to a polar perturbation. The dotted line is the diffusion
mode
$\mathfrak{w}_{I}=\mathfrak{q}^{2}$, which approaches the
quasinormal frequency in the hydrodynamic limit
$\mathfrak{w},\mathfrak{q}\ll 1$.}
\label{hidroeletro}}
%\end{figure}

As seen above, the hydrodynamic limit of perturbation equations comprises a
very special interval in the space of parameters $\mathfrak{w}$ and
$\mathfrak{q}$. Besides the physical relevance of this regime, it
corresponds to the very rare situations where analytical expressions
can be found for the quasinormal frequencies. In the great majority of
cases, numerical methods have to be employed in order to find
the complete dispersion relations
$\mathfrak{w}\times \mathfrak{q}$. In the sequence, the Horowitz-Hubeny
method \cite{horowitz1} is used to compute the dispersion relation
$\mathfrak{w}=-i\mathfrak{w}_{I}(\mathfrak{q})$ for the
electromagnetic hydrodynamic QNM, which in the limit of small wavenumbers
corresponds to the diffusion mode found above. The result is shown in Fig.
\ref{hidroeletro}. One sees the deviation of the exact dispersion relation
curve (solid line) from the hydrodynamic limit curve $\mathfrak{w}_{I} =
\mathfrak{q}^{2}$ (dotted line). Another interesting fact is that the
quasinormal frequency $\mathfrak{w}(\mathfrak{q})$ disappears
for $\mathfrak{q}$ larger than approximately $0.557$. This is
characteristic to all the electromagnetic purely damped modes,
as analyzed in the next section (see Table \ref{valoreslimites}).

\subsection{Dispersion relations for the non-hydrodynamic QNM}
\label{dispersion-eletro}

As mentioned earlier, one of the goals of the present analysis is to
obtain the electromagnetic quasinormal modes of plane-symmetric
AdS$_4$ spacetimes and to compare the present results with the
results of Ref. \cite{cardoso5}. As verified in the hydrodynamic limit
(Sect. \ref{hydro-eletro}), the QNM spectra calculated here may be quite
different from the spectra obtained in that work because
of the use of different fundamental variables, and therefore a more careful
search study on the dispersion relations of these modes is justified.

\subsubsection{Purely damped modes}
\label{purelydamped}

For small values of $\mathfrak{q}$, electromagnetic perturbations
of AdS black holes present a special set of
modes which are purely damped. These are not usual QNM since the real part
of the frequencies vanishes eliminating the oscillatory behavior of the
perturbations which is characteristic of QNM. Furthermore, the
frequencies $\mathfrak{w}(\mathfrak{q})$ of such modes cannot be, in
general, associated to hydrodynamic poles since most of the purely
damped modes have nonvanishing frequencies in limit as the wave\-number
$\mathfrak{q}$ goes to zero. To distinguish from the regular QNM, the purely
imaginary frequencies of both the axial and the polar perturbations shall be
labelled by a special quantum number, $n_s$, that assumes just integer
values, starting by $n_s=0$ for the hydrodynamic diffusion mode. As it was
shown above, the axial sector of electromagnetic perturbations does not
present quasinormal modes in the hydrodynamic limit, so that the set
of axial purely damped modes starts at $n_s=1$.
\FIGURE{
%\begin{figure}[t]
\centering\epsfig{file=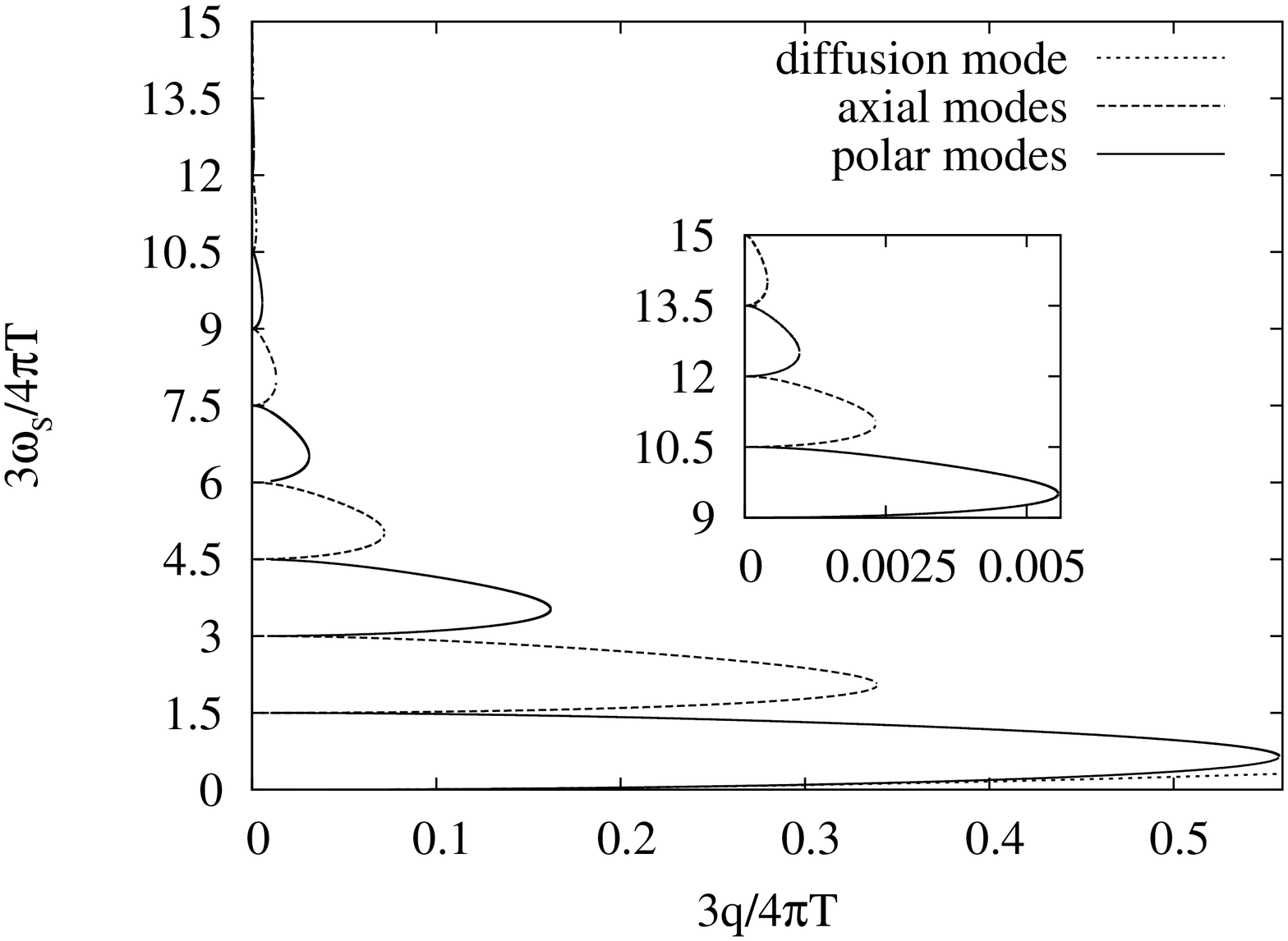, height=9.41cm,
width=6.0cm, angle=270}
\caption{The dispersion relations for polar (solid lines) and
axial (dashed lines) purely damped electromagnetic QNM. The dotted (lowest)
line is the diffusion mode $\mathfrak{w}_{s}=\mathfrak{q}^{2}$, which
approaches the $n_s =0$ quasinormal frequency in the hydrodynamic limit
$\mathfrak{w},\mathfrak{q}\ll 1$. The insert shows the
behavior of the dispersion relations for higher quasinormal
frequencies.}
\label{eletropuro}}
%\end{figure}

An interesting property of electromagnetic QNM of AdS black  holes in
$(3+1)$-di\-men\-sion\-al spacetimes has been recently discovered by
Herzog and collaborators \cite{herzog4}: The current-current
correlators are analytical functions at $\mathfrak{q}=0$, meaning
that there are no quasinormal frequencies for null wavenumber.
It can be shown that such a property is a consequence of
the well known duality relation between electric and magnetic
fields in vacuum. In fact, using the invariance of Maxwell equations under
the duality operation, electric field $\leftrightarrow$ magnetic field, and
the invariance of the correlation functions under rotations in the case of
null wavenumber (zero momentum), it was shown that the transverse and
longitudinal self-energies, $\Pi^{T}(\mathfrak{w},0)$ and
$\Pi^{L}(\mathfrak{w},0)$, are well behaved functions of the frequency for
all values of $\mathfrak{w}$.

On the other hand, as verified through the numerical results for purely
damped modes, there are quasinormal frequencies even for wavenumbers very
close to zero. In fact, as shown in Fig. \ref{eletropuro}, the small
wavenumber limit ($\mathfrak{q}=\epsilon$, with $\epsilon$ very small but
non-zero\footnote{In this specific case, the time of computation spent by
the numerical code based on the Horowitz-Hubeny method \cite{horowitz1}
written to find the quasinormal frequencies becomes very large as
$\epsilon$ approaches zero.}) of the corresponding purely imaginary
quasinormal frequencies, $\mathfrak{w}=-i\mathfrak{w}_{s}$, is given
approximately by
\begin{equation}
\mathfrak{w}_{s}=\frac{3}{2}n_{s}\qquad\Longrightarrow\qquad
\omega_{s}=\omega_{n_{s}}\equiv 2\pi T n_{s},
\end{equation}
where $\omega_{n_{s}}$ are the Matsubara frequencies of a generic quantum bosonic
system. Moreover, as it is also seen from Fig. \ref{eletropuro}, the
hydrodynamic pole $n_{s}=0$ is similar to other purely damped modes. As a
matter of fact, the only property that distinguishes a particular purely
damped mode from another is the behavior of these modes around
$\mathfrak{q}=0$: The only QNM satisfying the condition $\lim_{\mathfrak{q}
\rightarrow 0} \mathfrak{w}(\mathfrak{q}) =0$ is the hydrodynamic mode.

\TABLE{
%\begin{table}[t]
\begin{tabular}{cccccc}
\hline\hline
\multicolumn{3}{c}{Polar} &
\multicolumn{3}{c}{Axial}\\
\cline{1-6}
$n_{s}$ & $\mathfrak{q}_{\mbox{\scriptsize{lim}}}\times 10^{3}$
& $\mathfrak{w}_{\mbox{\scriptsize{lim}}}$ (interval) &
$\quad n_{s}$ & $\mathfrak{q}_{\mbox{\scriptsize{lim}}}\times
10^{3}$ & $\mathfrak{w}_{\mbox{\scriptsize{lim}}}$ (interval)\\
\hline
(0,1) & $557.319\;$ &  $[0.648111$, $0.648429]$ &
$\quad$(1,2) & $339.330$  &  $[2.04771$, $2.04811]$ \\
(2,3) & $162.034\;$ &  $[3.52507$, $3.52562]$ &
$\quad$(3,4) & $71.8726$  &  $[5.01701$, $5.01788]$ \\
(4,5) & $31.0102\;$ &  $[6.51286$, $6.51384]$ &
$\quad$(5,6) & $13.1892$  &  $[8.00994$, $8.01169]$ \\
(6,7) & $5.55917\;$ &  $[9.50785$, $9.51033]$ &
$\quad$(7,8) & $2.32839$ &  $[11.0057$, $11.0100]$ \\
(8,9) & $0.970660\;$ &  $[12.5041$, $12.5097]$ &
$\quad$(9,10) & $0.403180$ &  $[14.0009$, $14.0114]$ \\
\hline\hline
\end{tabular}
\centering
\caption{Approximate limiting values of frequencies
$\mathfrak{w}=-i\mathfrak{w}_{\mbox{\scriptsize{lim}}}$
and wavenumbers $\mathfrak{q}_{\mbox{\scriptsize{lim}}}$
for purely damped electromagnetic modes. The brackets
indicate that the actual limiting values lie between
the two indicated endpoints in each case.}
\label{valoreslimites}}
%\end{table}

At the opposite side of the quasinormal spectrum, i.e., for larger values of
$\mathfrak{q}$, there are saturation points at a maximum wavenumber value,
$\mathfrak{q}_{\mbox{\scriptsize{lim}}}$, beyond which the specific mode
disappears (See, however, Figs. \ref{eletroaxial} and \ref{eletropolar}). This
seems to happen for everyone of the modes, with
$\mathfrak{q}_{\mbox{\scriptsize{lim}}}$ decreasing for higher overtones (cf.
Fig. \ref{eletropuro}). The curves representing dispersion relations
associated to two different but contiguous modes meet each other exactly at
the saturation point, i.e., the two dispersion relation curves coincide at
that point. The axial modes group in pairs according to the relation
$n_{s}=\{(1,2),(3,4),(5,6),...\}$, while the polar modes are paired as
$n_{s}=\{(0,1),(2,3),(4,5),...\}$. The approximate limiting values
$\mathfrak{w}_{\mbox{\scriptsize{lim}}}$ and
$\mathfrak{q}_{\mbox{\scriptsize{lim}}}$ for $n_{s}=0,1,2,...,10$  are shown
in Table \ref{valoreslimites}.

The  existence of a meeting point between two contiguous dispersion relation
curves suggests that for the special wavenumber values
$\mathfrak{q}=\mathfrak{q}_{\mbox{\scriptsize{lim}}}$ the corresponding
quasinormal frequencies
$\mathfrak{w}=-i\mathfrak{w}_{\mbox{\scriptsize{lim}}}$ represent double poles
of the CFT current-current correlation functions. A strong support to such a
conclusion comes from the behavior of the connection coefficients
$\mathcal{A}_{(j)}(\mathfrak{w}, \mathfrak{q})$ as a function of
$\mathfrak{w}=\mathfrak{w}_{R}-i\mathfrak{w}_{I}$ for small values of
$\mathfrak{q}$ and $\mathfrak{w}_{R}=0$. In Fig. \ref{funcaocompleta} it is
shown the profile of $\mathcal{A}_{(x)}(\mathfrak{w},\mathfrak{q})$ for
$\mathfrak{q}=\mathfrak{q}_{\mbox{\scriptsize{lim}}} \simeq 0.0132$ and
$0<\mathfrak{w}_{I}<9$.  For this wavenumber, there are six purely imaginary
quasinormal frequencies encompassing the $n_{s}=1$ to $n_{s}=6$ axial QNM,
which by definition are the points where
$\mathcal{A}_{(x)}(\mathfrak{w},\mathfrak{q})=0$. In particular, for $n_{s}=5$
and $n_{s}=6$, the zeros of $\mathcal{A}_{(x)} (\mathfrak{w},\mathfrak{q})$
coincide, indicating that the corresponding quasinormal frequencies are
identical (see also Fig. \ref{eletropuro}).

The multiplicity of specific quasinormal frequencies as poles of
the correlation functions may also be verified through the derivatives of
$\mathcal{A}_{(j)}(\mathfrak{w},\mathfrak{q})$ with respect to 
$\mathfrak{w}$. This was done numerically, by taking fixed values of 
$\mathfrak{q}$ and letting $\mathfrak{q} \rightarrow
\mathfrak{q}_{\mbox{\scriptsize{lim}}}$. It was found that the first
derivative of $\mathcal{A}_{(j)}(\mathfrak{w},\mathfrak{q})$ in relation
to $\mathfrak{w}$ vanishes when $\mathfrak{w}=
-i\mathfrak{w}_{\mbox{\scriptsize{lim}}}$ and
 $\mathfrak{q}=\mathfrak{q}_{\mbox{\scriptsize{lim}}}$, but it
is not zero for other values of the wavenumber. 
This result proves that $\mathfrak{w}=-i\mathfrak{w}_{\mbox{\scriptsize{lim}}}$
corresponds to, at least, a double zero of $\mathcal{A}_{(j)}(\mathfrak{w},
\mathfrak{q})$, and consequently, to a double pole of the
corresponding current-current correlation functions.

\FIGURE{
%\begin{figure}[t]
\centering\epsfig{file=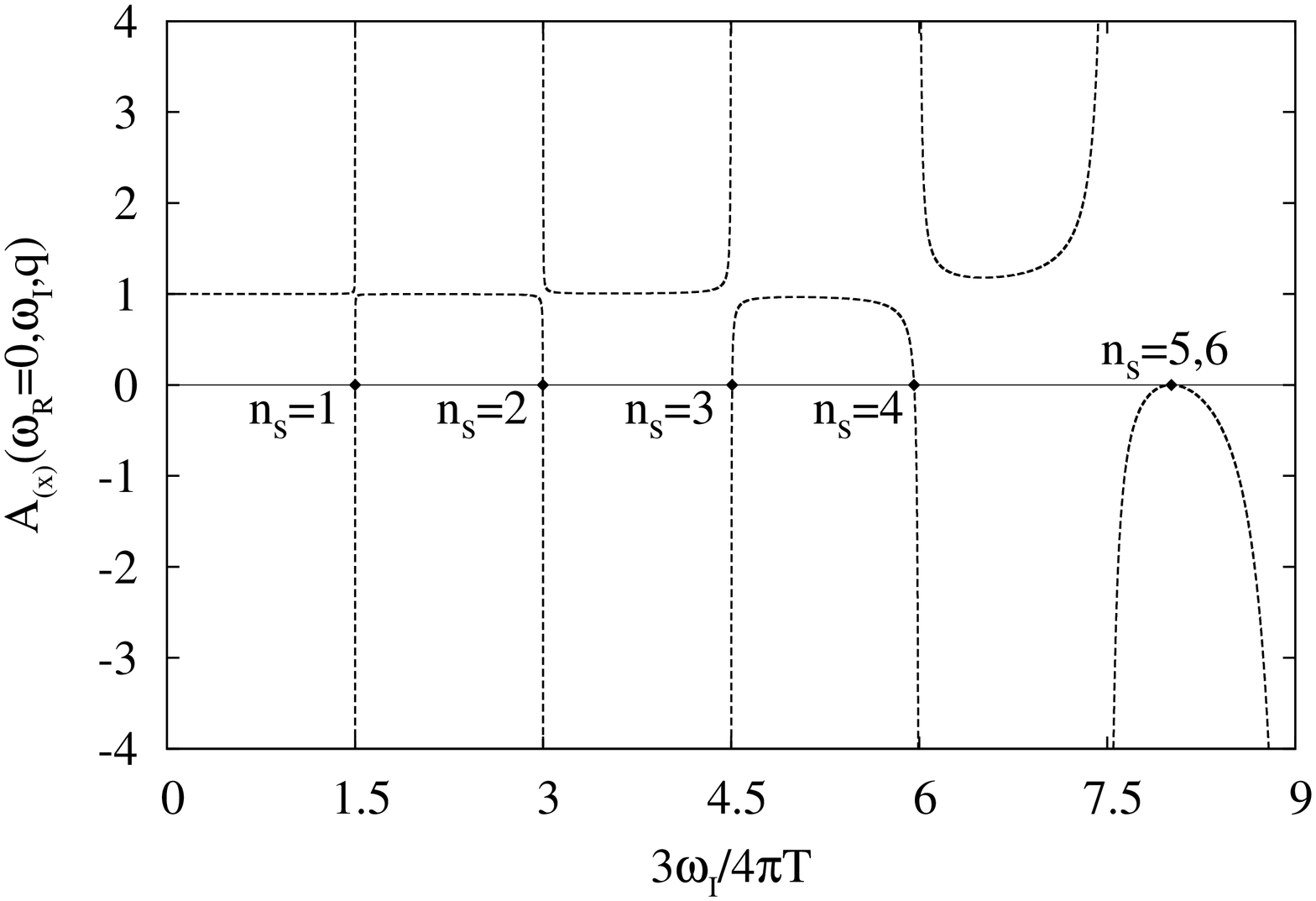, height=9.91cm,
width=7.0cm, angle=270} \caption{The connection coefficient
$\mathcal{A}_{(x)}(\mathfrak{w}, \mathfrak{q})$ for $\mathfrak{w}_{R}=0$,
$\mathfrak{w}_{I}=(0,9)$ and 
$\mathfrak{q}=\mathfrak{q}_{\mbox{\scriptsize{lim}}}\simeq 0.0132$. The points
represent the $n_{s}=1,2,..,6$ purely imaginary quasinormal frequencies
associated to the axial electromagnetic perturbations. Note the coincidence of
the points corresponding to $n_{s}=5$ and $n_{s}=6$, indicating a possible
doubleness of the related quasinormal frequency.}
\label{funcaocompleta}}
%\end{figure}

\subsubsection{Ordinary quasinormal modes}

The electromagnetic perturbations of AdS black holes present also a family of
regular (ordinary) quasinormal modes whose frequencies have nonzero real and
imaginary parts. The numerical results for the quasinormal frequencies of the
first five regular modes are shown respectively in Figs.
\ref{eletroaxial} and \ref{eletropolar} for axial and polar fluctuations.

The form of the dispersion relations $\mathfrak{w}_{R}(\mathfrak{q})$
and $\mathfrak{w}_{I}(\mathfrak{q})$ indicates a connection between
the electromagnetic ordinary QNM and the family of purely damped modes
discussed in the last section. As shown in Figs. \ref{eletroaxial} and
\ref{eletropolar}, each regular quasinormal frequency only appears for
$\mathfrak{q}$ larger than a minimum wavenumber value, which (to a good
approximation) coincides with the limiting value of the corresponding pair
of purely damped modes, $\mathfrak{q}_{\mbox{\scriptsize{lim}}}$.
That is to say, all the dispersion relations, $\mathfrak{w} (\mathfrak{q})$,
for the ordinary QNM begin at the points $(\mathfrak{w}_{\mbox{\scriptsize{lim}}},
\mathfrak{q}_{\mbox{\scriptsize{lim}}})$, with the real parts starting from
zero value, $\displaystyle{ \mathfrak{w}_{R}(\mathfrak{q}\rightarrow
\mathfrak{q}^+_{\mbox{\scriptsize{lim}}})=0}$, while the imaginary parts
start at $\mathfrak{w}_{I}(\mathfrak{q}\rightarrow\mathfrak{q}^+
_{\mbox{\scriptsize{lim}}})=\mathfrak{w}_{\mbox{\scriptsize{lim}}}$.
For higher wavenumber values, the ordinary electromagnetic modes
show a sequence of quasinormal frequencies whose imaginary parts grow with
the principal quantum number $n$. In this respect, Figs. \ref{eletroaxial}
and \ref{eletropolar} show a simiral behavior for the real parts of the
quasinormal electromagnetic frequencies.

\FIGURE{
%\begin{figure}[t]
\centering\epsfig{file=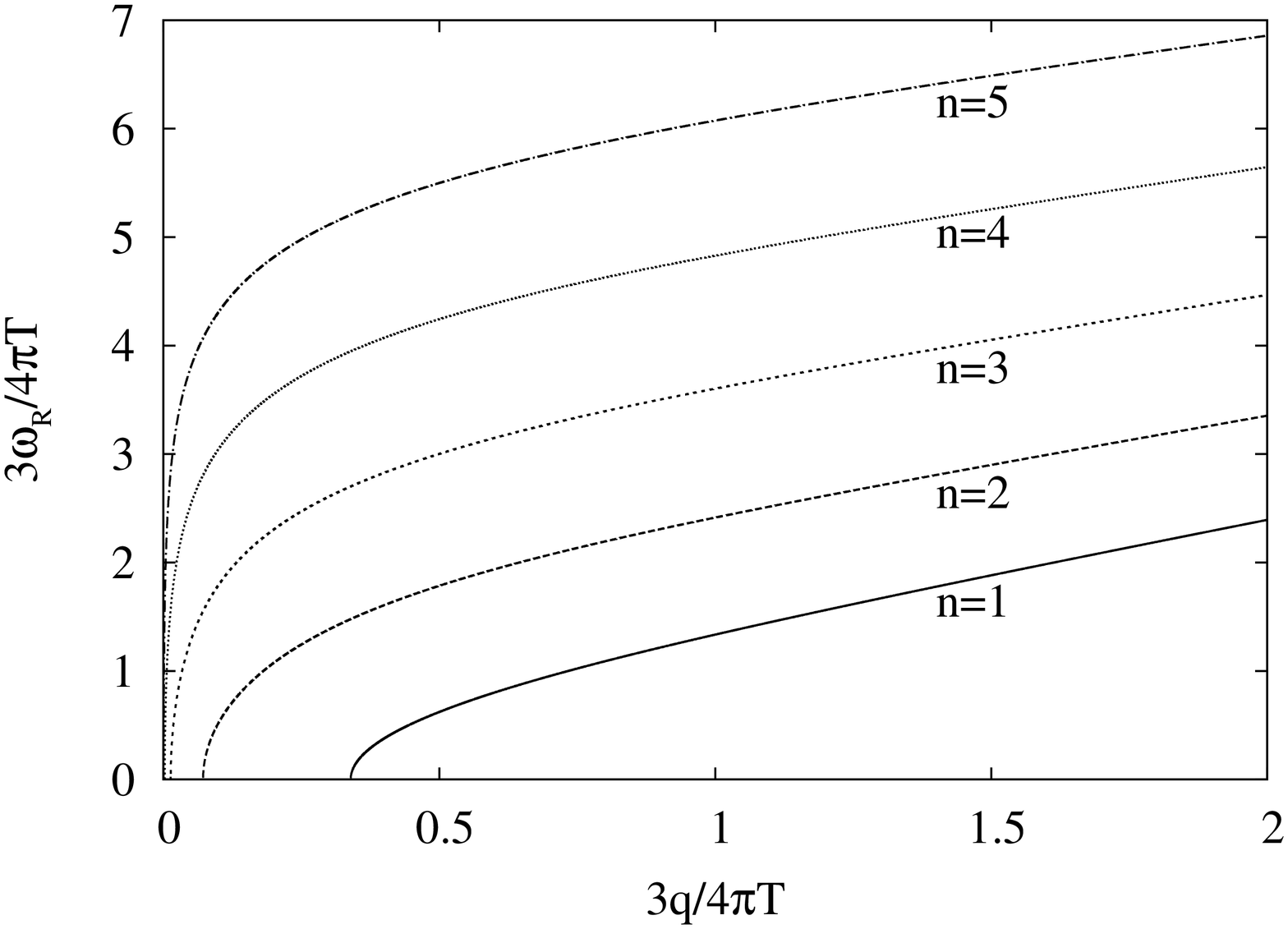, height=7.137cm,
width=4.968cm, angle=270}
\centering\epsfig{file=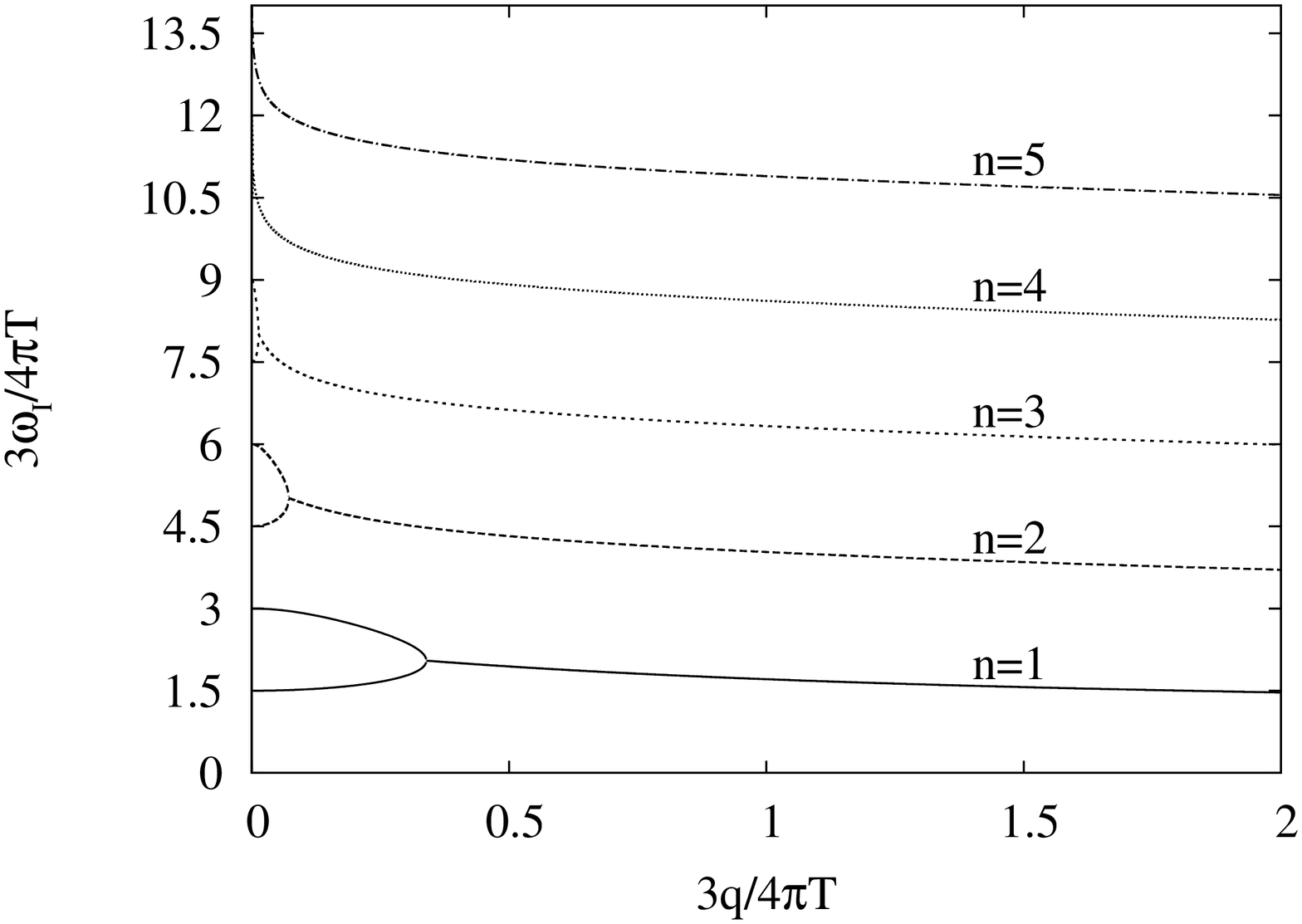, height=7.137cm,
width=4.968cm, angle=270}
\caption{The real (left) and imaginary (right) parts of the frequencies
$\mathfrak{w}=3\omega/4\pi T$
for the first five ordinary axial electromagnetic modes
as a function of the normalized wavenumber $\mathfrak{q}=3q/4\pi T$.
The quantum number $n$ arranges the regular polar QNM in increasing
order of values of $\mathfrak{w}_{I}$. In the right it is also shown
the frequencies $\mathfrak{w}_{s}=3\omega_{s}/4\pi T$ associated to
the axial electromagnetic purely damped QNM.}
\label{eletroaxial}}
%\end{figure}
\FIGURE{
%\begin{figure}[t]
\centering\epsfig{file=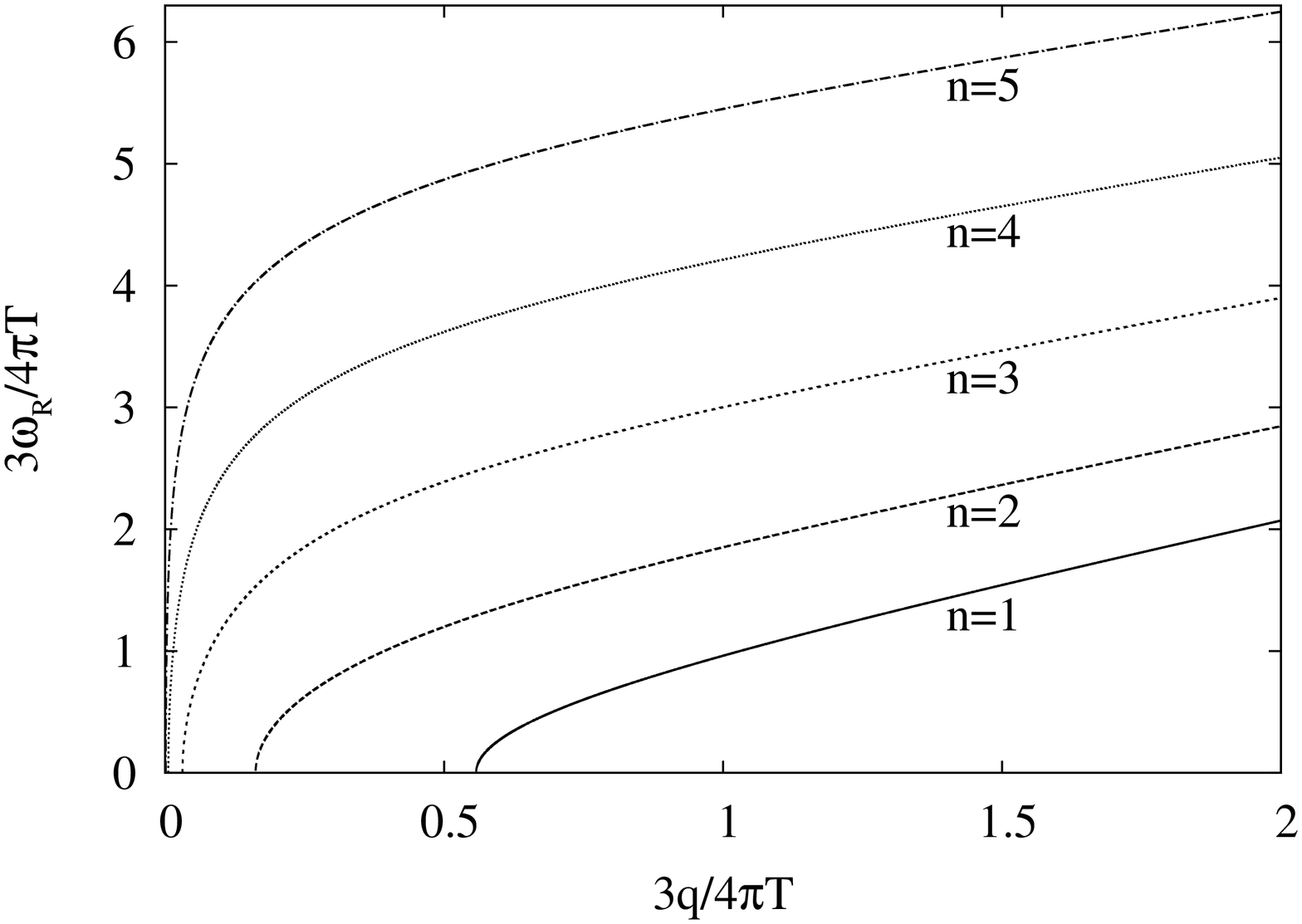, height=7.137cm,
width=4.968cm, angle=270}
\centering\epsfig{file=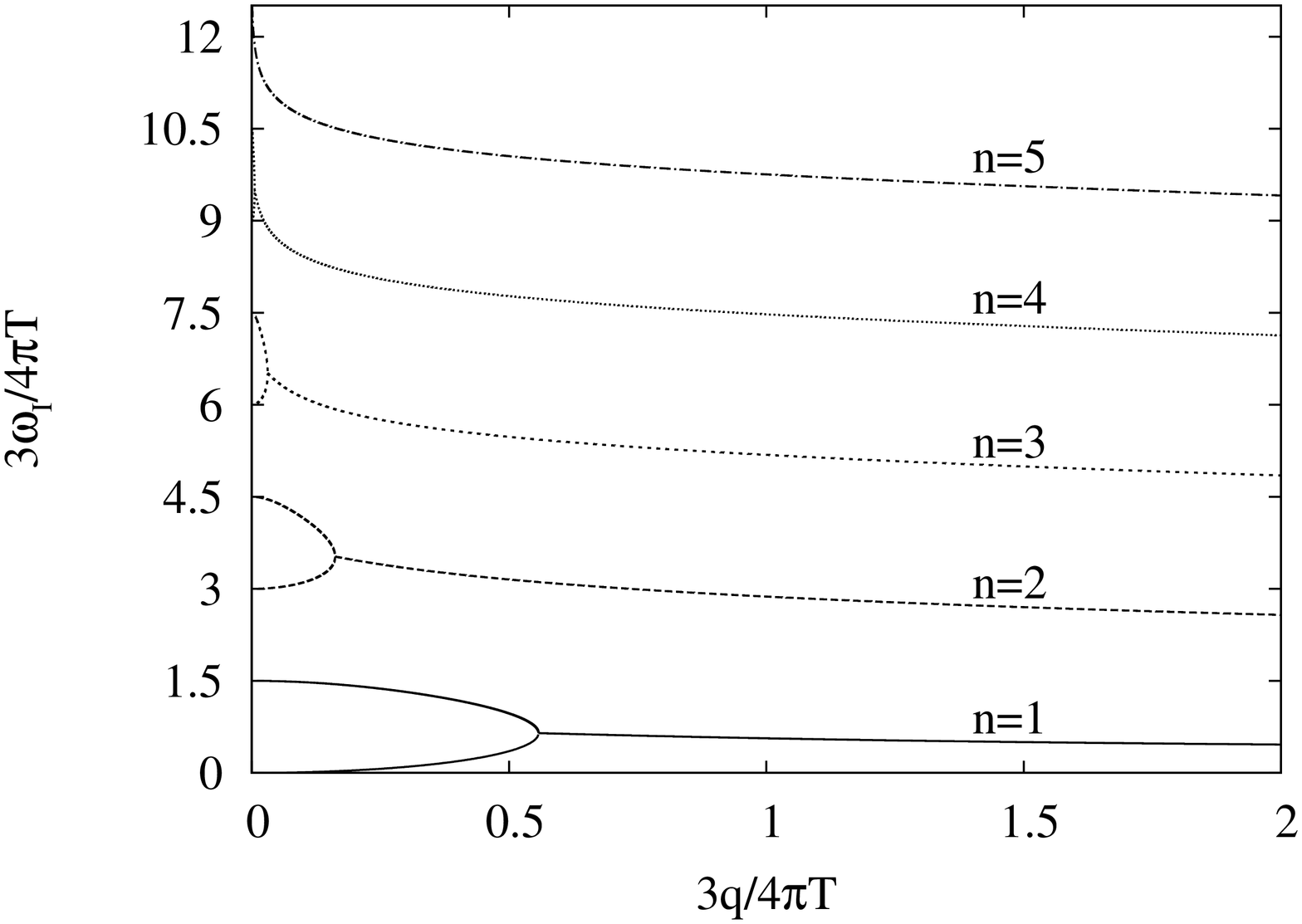, height=7.137cm,
width=4.968cm, angle=270}
\caption{The real (left) and imaginary (right) parts of the frequencies
$\mathfrak{w}=3\omega/4\pi T$ for the first five ordinary polar
electromagnetic modes as a function of the normalized wavenumber
$\mathfrak{q}=3q/4\pi T$. As for the axial modes, the quantum number $n$
arranges the regular QNM from lower to higher values of $\mathfrak{w}_{I}$.
In the right it is also shown the frequencies $\mathfrak{w}_{s}=
3\omega_{s}/4\pi T$ associated to the polar electromagnetic purely
damped QNM.}
\label{eletropolar}}
\FIGURE{
%\begin{figure}[t]
\centering\epsfig{file=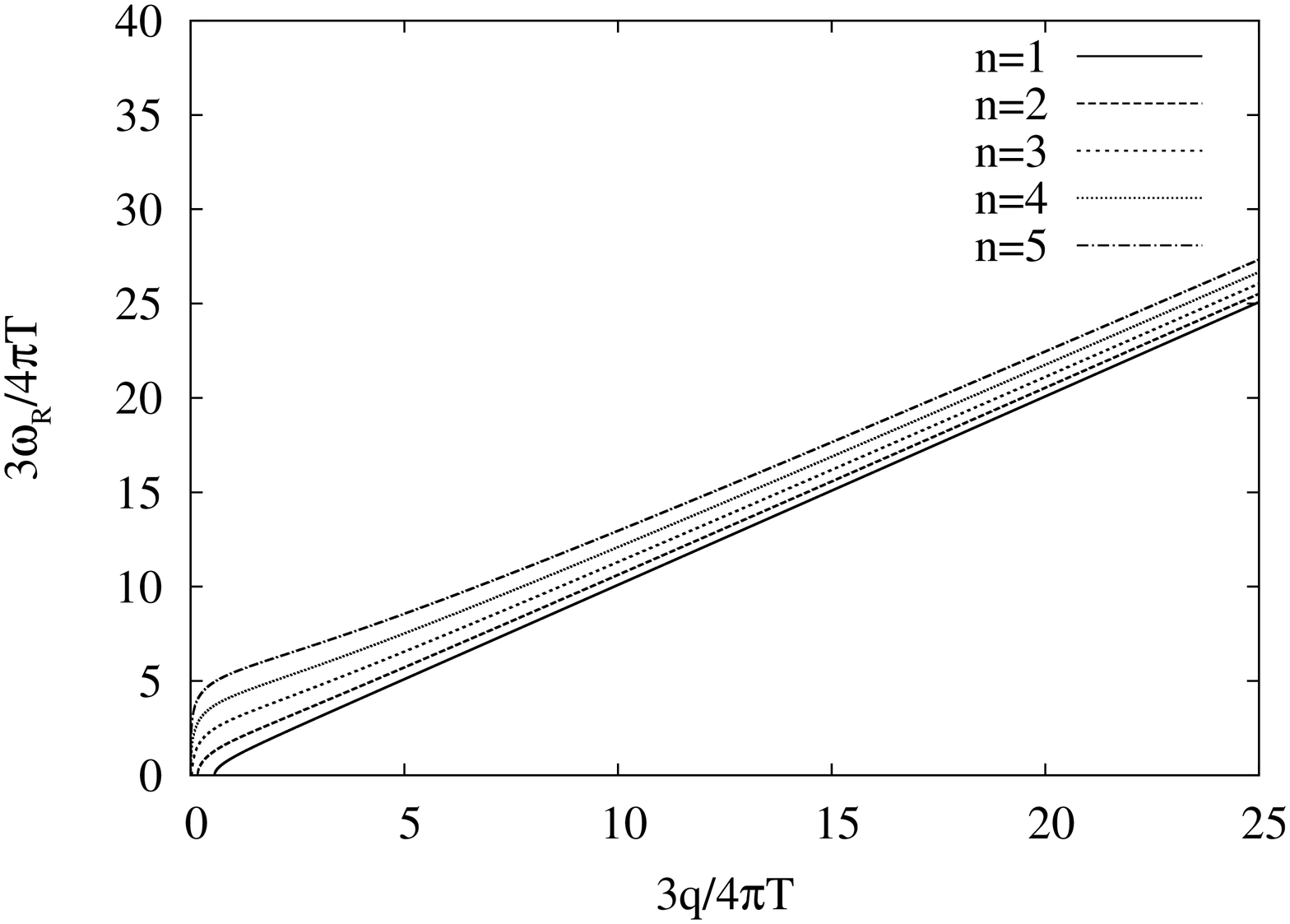, height=7.337cm,
width=5.668cm, angle=270}
\centering\epsfig{file=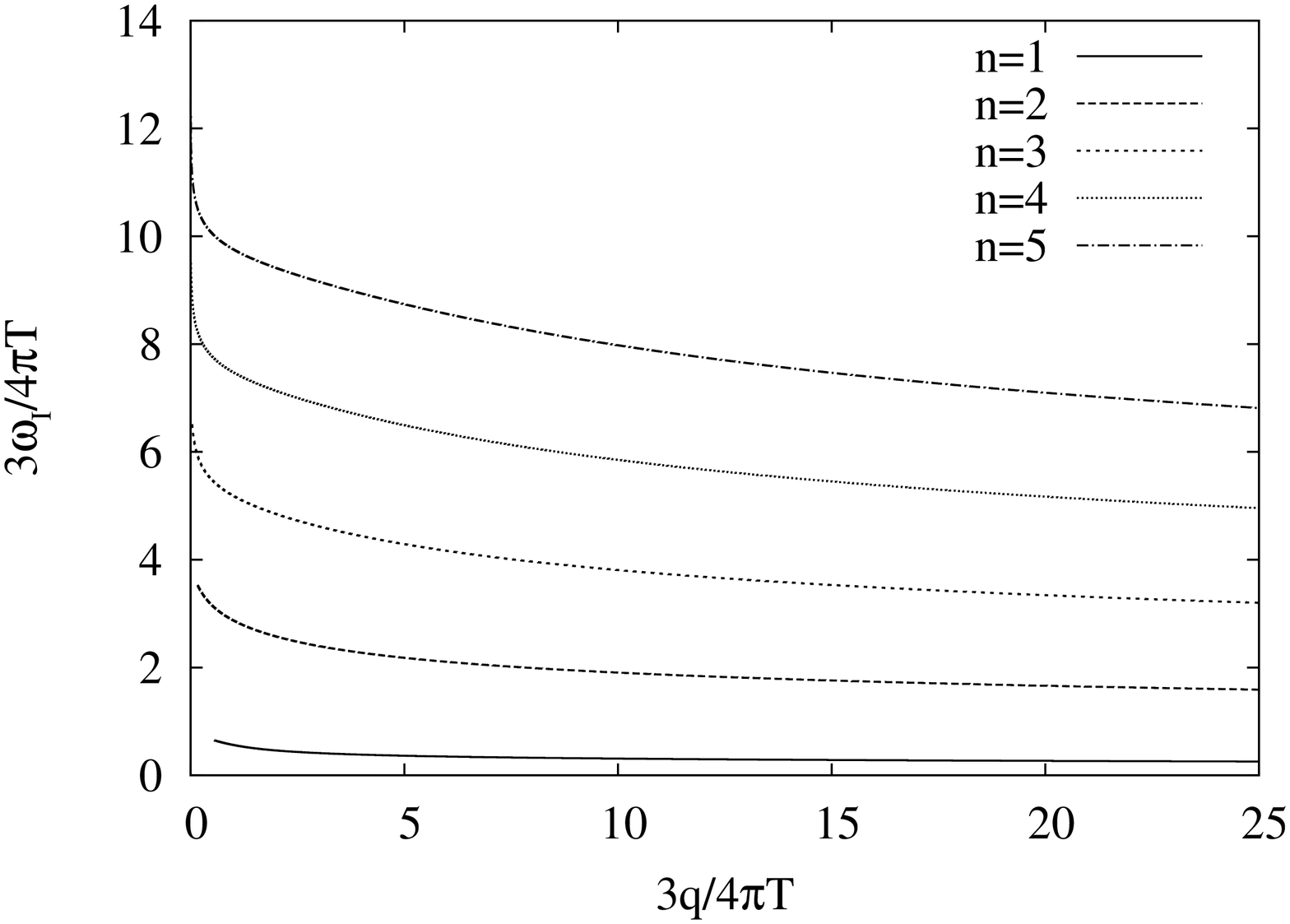, height=7.337cm,
width=5.668cm, angle=270}
\caption{Graphs of the dispersion relations of the first five regular polar
electromagnetic QNM for large values of $\mathfrak{q}$. The figure on the
left hand side shows the real part of the frequency, $\mathfrak{w}_{R}$,
while the graph on the right hand side is for the imaginary part,
$\mathfrak{w}_{I}$.}
\label{eletroexpandido}}
%\end{figure}

The quasinormal frequencies found here show that AdS black holes are not
good oscillators. As it is well known, an interesting way of measuring the
quality of an oscillator is by means of its quality factor
$Q=\mathfrak{w}_{R}/2\mathfrak{w}_{I}$. In general, in the region of small
wavenumbers, the electromagnetic QNM have very small quality factors, $Q\ll
1$. For instance, taking $\mathfrak{q}=0.557319$ and considering the
fundamental polar mode one obtains $Q=7.24\times 10^{-5}$, a quality factor
typical to highly damped oscillatory systems. On the other hand, quality
factors of the order of unity are found for large wavenumber values, such as
$Q=1$ for $\mathfrak{q}$ around $1.11$, and $Q= 85.9$ for
$\mathfrak{q}=40$.

From the holographic field theory point of view, the real part of the
frequencies may be interpreted as quasiparticle excitation energies in the
dual plasma defined at the conformal boundary of the AdS spacetime. However,
such an interpretation only
makes sense for excitations, or quasinormal modes, with
large quality factors. In fact, according to Heisenberg uncertainty
principle, the uncertainty in the energy of a quasiparticle is of the
order of $\omega_{I}$ (in units of $\hbar$). Hence, quality factors smaller
than unity imply in energy uncertainties larger than the energies of
the quasiparticles themselves, making the interpretation of energy
excitations as quasiparticles meaningless. 

In the $Q\gg 1$ regime, where the quasiparticle interpretation is feasible,
Fig. \ref{eletroexpandido} shows that the dispersion relations of the
ordinary electromagnetic QNM frequencies have real parts approaching
straight lines of the form $\mathfrak{w}_{R}=\mathfrak{q}+\mathfrak{b}_{n}$,
where $\mathfrak{b}_n$ depends only on the specific mode $n$. This means the
functions $\omega_{R}(q)$ approach the usual energy-momentum relation associated
to a zero rest mass particle, $\omega_{R}=q$, as $T\rightarrow 0$. Furthermore,
the characteristic damping time ($\tau=1/\omega_{I}$) of the
electromagnetic fluctuations diverges in the limit
$\mathfrak{q}\rightarrow\infty$, i.e., the functions $\omega_I(q)$ tend to
zero for large wavenumbers. All of the above results are consistent with
the expected properties of poles of correlation functions in quantum field
theories at zero temperature \cite{herzog4}.

\section{Gravitational quasinormal modes}
\label{grav-qnm}

Even though gravitational perturbations of plane-symmetric black holes in
$(3+1)$-dimensional AdS spacetimes have been analyzed in some extent
\cite{herzog2, herzog3, cardoso5, miranda1, miranda2}, the
quasi\-nor\-mal-mode
dispersion relations, with the use of the KS variables, were not found yet,
and hence the comparison with the spectra obtained by using the RWZ
gauge-invariant variables was not performed. This is done next.

\subsection{Fundamental equations for gravitational fluctuations}
\label{flut-grav}

As usual, gravitational perturbations are described here in terms of linear
metric fluctuations, which means the metric for the perturbed spacetime is
written as $g_{\ss{MN}}=g^{\ss{0}}_{\ss{MN}} + h_{\ss{MN}}$, where
$h_{\ss{MN}}$ is considered as a perturbation in the background metric
$g^{\ss{0}}_{\ss{MN}}$, given by Eq. \eqref{background}. Among the variety
of possible gauge choices in studying metric perturbations, an interesting
choice is the so-called radial gauge, in which the coordinate system is
chosen in such a way that $h_{rt}=h_{rx}=h_{ry}=h_{rr}=0$. Since one of
the aims here is to investigate the relation among the perturbations of
plane-symmetric AdS$_4$ black holes and the correlation functions in the dual
CFT, it is convenient to use the radial gauge formalism to study the metric
fluctuations. A brief description of this formalism is given in the
following.

As it was done in the previous section when studying the electromagnetic
perturbations, the isometries of spacetime \eqref{background} allow Fourier
transforming coordinates $t$, $x$ and $y$, and writing metric fluctuations
$h_{\ss{MN}}$ as
\begin{equation}
h_{\ss{MN}}(t,x,y,r)= \frac{1}{(2\pi)^{3}} \int \!{d\omega\, dk_{x}\,
dk_{y}\,} e^{-i\omega t+ik_{x}x+ik_{y}y}
\widetilde{h}_{\ss{MN}}(\omega,k_{x},k_{y},r). 
\label{gravfourier}
\end{equation}
Again the wave three-vector may be chosen as $k_\mu=(-\omega,0,q)$,
and hence metric perturbations can be split into two disjoint sets. Namely,
the axial (transverse) sector of gravitational perturbations is
characterized by the quantities $h_{tx}$ and $h_{yx}$, and the polar
(longitudinal) sector of perturbations is composed by $h_{tt}$,
$h_{xx}$, $h_{yy}$, and $h_{ty}$.

\subsubsection{Axial perturbations in the radial gauge}

Now one needs to find evolution equations for the axial (transverse)
gravitational perturbations in the radial gauge, that are composed by the
metric fluctuations $h_{tx}$ and $h_{yx}$. The task is carried out following
the usual procedure of the theory of linear perturbations. The linearized
Einstein equations corresponding to the axial sector of gravitational
perturbations yield a set of three coupled differential equations for
$h_{tx}$ and $h_{yx}$ \cite{herzog2}. Of course, such equations are not
independent from each other, and one of them may be eliminated as a
combination of the other two. A resulting system of linearly independent
equations (after Fourier transforming them) which is interesting for the
present study is the following:
\begin{gather}
H_{tx}^{'}+\frac{\mathfrak{q}\mathfrak{h}}{\mathfrak{w}}
H_{yx}^{'}=0,\label{axial1}\\
H_{tx}^{''}-\frac{2}{u}H_{tx}^{'}-\frac{\mathfrak{q}}{\mathfrak{h}}
\left(\mathfrak{w}H_{yx}+\mathfrak{q} H_{tx}\right)=0, 
\label{axial2}
\end{gather}
where $H_{tx}$ and $H_{yx}$ are defined by
\begin{equation}
H_{tx}=\frac{L^{2}}{r^{2}}\,h_{tx},\qquad\qquad
H_{yx}=\frac{L^{2}}{r^{2}}\,h_{yx}.
\label{axialh-ij}
\end{equation}

As in the case of electromagnetic perturbations, a mandatory condition that a
candidate for fundamental variable must satisfy is being gauge invariant,
which for metric perturbations means the candidate has to be invariant under
infinitesimal coordinate transformations. Inspired once again in the work by
Kovtun and Starinets \cite{kovtun1}, among the different combinations of
axial functions $H_{tx}$ and $H_{yx}$ which furnish gauge-invariant
quantities, one takes
\begin{equation}
Z_{1}=i\left(qH_{tx}+\omega H_{yx}\right)
\label{mestreaxial}
\end{equation}
as the fundamental gauge-invariant function of
the axial gravitational perturbations.

Decoupling the system of differential equations \eqref{axial1} and
\eqref{axial2} in terms of the fundamental variable $Z_{1}$,
it results the solely second-order differential equation
\begin{equation}
Z_{1}^{''}-\frac{2(\mathfrak{w}^{2}-\mathfrak{q}^{2}
\mathfrak{h})\mathfrak{h}-u\mathfrak{h}^{'}
\mathfrak{w}^{2}}{u\mathfrak{h}(\mathfrak{w}^{2}-\mathfrak{q}^{2}
\mathfrak{h})}Z_{1}^{'}
+\frac{\mathfrak{w}^{2}-\mathfrak{q}^{2}
\mathfrak{h}}{\mathfrak{h}^{2}}Z_{1}=0.
\label{yeqaxial}
\end{equation}
Solutions to this equation satisfying the QNM boundary conditions are
studied below.

\subsubsection{Polar perturbations in the radial gauge}

The polar (longitudinal) sector of gravitational perturbations in the radial
gauge is described by metric  fluctuations $h_{tt}$, $h_{xx}$, $h_{yy}$ and
$h_{ty}$. These components of the metric perturbation tensor are used to
define new quantities 
\begin{equation}
H_{tt}=\frac{1}{f}\,h_{tt},\qquad
H_{xx}=\frac{L^{2}}{r^{2}}\,h_{xx},\qquad
H_{yy}=\frac{L^{2}}{r^{2}}\,h_{yy},\qquad
H_{ty}=\frac{L^{2}}{r^{2}}\,h_{ty},
\label{polarh-ij}
\end{equation}
which are more appropriate to deal with during calculations to obtain
perturbation equations. Hence, the polar components of linearized Einstein
equations furnish a set of seven coupled equations for the variables defined
in Eqs. \eqref{polarh-ij}. Only four of such a set are independent equations
and, among the possible choices, the more interesting for the present work
are
\begin{align}
H_{ty}^{'}=&\frac{2u\mathfrak{w}\mathfrak{q}}{b(u)} \left(H_{xx}
-H_{tt}\right)+ \frac{ua(u)}{2\mathfrak{q}\mathfrak{h}b(u)}
\left(\mathfrak{w}H_{xx} +\mathfrak{w}H_{yy}+ 2\mathfrak{q}H_{ty}\right)
-\frac{4\mathfrak{w}\mathfrak{h}}{\mathfrak{q} b(u)}H_{tt}^{'},
\label{polar1}\\ H_{xx}^{'}=&\frac{\mathfrak{w}ua(u)}{2\mathfrak{q}^{2}
\mathfrak{h}^{2}b(u)} \left(\mathfrak{w}H_{xx} +\mathfrak{w}H_{yy}
+2\mathfrak{q}H_{ty}\right) -\frac{c(u)}{\mathfrak{q}^{2}b(u)}H_{tt}^{'}+
\frac{2u\mathfrak{w}^{2}} {\mathfrak{h}b(u)} H_{xx}+\frac{ua(u)}{2
\mathfrak{h}b(u)}H_{tt}, \label{polar2}\\ \notag\\[-0.6cm]
H_{yy}'=&\frac{2u}{\mathfrak{h}b(u)}\left[\mathfrak{w}^{2} H_{xx}
+\mathfrak{w}^{2}H_{yy} +\mathfrak{q}^{2}\mathfrak{h} \left(H_{tt}
-H_{xx}\right) +2\mathfrak{q}\mathfrak{w} H_{ty}\right]
+\left(\frac{4\mathfrak{h}}{b(u)} +\frac{c(u)}{\mathfrak{q}^{2}
b(u)}\right)H_{tt}^{'} \nonumber\\ -&\frac{u}{2\mathfrak{h}b(u)}
\left[4\mathfrak{w}^{2} H_{xx} +c(u)H_{tt}\right]
-\frac{\mathfrak{w}uc(u)}{2\mathfrak{q}^{2} \mathfrak{h}^{2}b(u)}
\left(\mathfrak{w}H_{xx} +\mathfrak{w}H_{yy} +2\mathfrak{q}H_{ty}\right),
\label{polar3}\\ \notag\\[-0.6cm]
\!\!\!H_{tt}^{''}=&\frac{2\mathfrak{w}^{2}}{\mathfrak{h}b(u)} \left(H_{xx}
+H_{yy}\right) +\frac{2\mathfrak{q}}{\mathfrak{h}b(u)}
\left(2\mathfrak{w}H_{ty} +\mathfrak{q}\mathfrak{h} H_{tt}\right)
+\frac{(2+u^{3})}{2u\mathfrak{h}b(u)} \left[\mathfrak{q}^{2}H_{xx}
+(8+u^3)H_{tt}^{'}\right], \label{polar4}
\end{align}
where, as above, the primes denote derivatives with respect to the variable
$u=r_0/r$, and coefficients $a(u)= 3u^{4}-12u-4\mathfrak{w}^{2}$,
$b(u)=\mathfrak{h}+3$, and $c(u)=4\mathfrak{w}^{2}-\mathfrak{q}^{2}b(u)$
were introduced to simplify notation.

Finally, a gauge-invariant function $Z_2$ is built as a particular
combination of the metric perturbations 
\begin{equation}
Z_2=4\omega qH_{ty}+2\omega^{2}H_{yy} +\left[q^{2}(3-\mathfrak{h})
-2\omega^{2}\right] H_{xx}+2q^{2}\mathfrak{h}H_{tt},
\label{mestrepolar}
\end{equation}
for which, uncoupling the equations of motion \eqref{polar1}-\eqref{polar4},
it is found the following second-order differential equation
\begin{equation}
Z_{2}^{''}-\frac{4\mathfrak{w}^{2}(2+u^{3})+\mathfrak{q}^{2}
d(u)}{u\mathfrak{h}c(u)}Z_{2}^{'} +\frac{4\mathfrak{w}^{4}
+\mathfrak{q}^{4}\mathfrak{h}b(u) -\mathfrak{q}^{2}e(u)}
{\mathfrak{h}^{2}c(u)}Z_{2}= 0,
\label{yeqpolar}
\end{equation}
where $d(u)=4u^{3}-5u^{6}-8$ and
$e(u)=9u^{4}\mathfrak{h}+\mathfrak{w}^{2}(8-5u^{3})$.
Equations \eqref{yeqaxial} and \eqref{yeqpolar} are the fundamental
equations that are going to be used in the next sections to compute
the QNM spectra associated to gravitational perturbations
of plane AdS$_4$ black holes.

\subsection{Stress-energy tensor correlation functions}
\label{green-gravit}

For the gravitational perturbations, the AdS/CFT correspondence
establishes a relation among the solutions of Eqs. \eqref{yeqaxial} and
\eqref{yeqpolar} and the stress-energy tensor of the dual CFT. From this
relation, the stress-energy tensor correlators can be determined, and in 
order to do that the explicit form of the fields in the bulk AdS spacetime
has to be known. More precisely, in order to impose the ingoing-wave
condition at the horizon, and to map AdS to CFT quantities at the boundary
of the spacetime, the asymptotic form of the metric perturbation
functions close to the horizon and at the boundary are necessary.

In the horizon neighborhood ($u\approx 1$), the gravitational
gauge-invariant variables $Z_1$ and $Z_2$ have a similar behavior as the
electric field components (see Sect. \ref{green-electro}), viz,
$Z_{1,2}\sim\mathfrak{h}^{\pm i\mathfrak{w}/3}$. As in the electromagnetic
case, to compute the retarded Green functions, one has to choose
the solutions corresponding to the negative imaginary power,
$Z_{1,2}\sim\mathfrak{h}^{-i\mathfrak{w}/3}$. On the other hand,
at the conformal boundary of the AdS spacetime, the metric fluctuations
are such that
\begin{gather}
Z_{1}=\mathcal{A}_{(1)}(\mathfrak{w},\mathfrak{q})
+...\;+\mathcal{B}_{(1)}(\mathfrak{w},\mathfrak{q})u^{3}+...\, ,
\label{assintgrav1}\\
Z_{2}=\mathcal{A}_{(2)}(\mathfrak{w},\mathfrak{q})+...\;+
\mathcal{B}_{(2)}(\mathfrak{w},\mathfrak{q})u^{3}+...\, ,
\label{assintgrav2}
\end{gather}
where ellipses denote higher powers of $u$, and  quantities
$\mathcal{A}_{(1)}(\mathfrak{w},\mathfrak{q})$ and 
$\mathcal{B}_{(1)}(\mathfrak{w},\mathfrak{q})$, and 
$\mathcal{A}_{(2)}(\mathfrak{w}, \mathfrak{q})$ and 
$\mathcal{B}_{(2)}(\mathfrak{w},\mathfrak{q})$ 
are the connection coefficients related to the differential equations 
\eqref{yeqaxial} and \eqref{yeqpolar}, respectively.

For the remaining of this section, as usual, the gravitational perturbations
are split into axial and polar sectors and the analysis of the corresponding
actions, coming from Eq. \eqref{acaocompleta}, are performed separately for
both of the perturbation types.

\subsubsection{Axial sector}

It is well known that in the calculation of two-point correlation functions
from the gravitational action only quadratic terms in metric perturbations
need to be considered. Moreover, according to the Lorentzian AdS/CFT
prescription \cite{son1} (see also Ref. \cite{policastro1}), in order to
obtain the CFT retarded correlators the relevant terms are the quadratic
terms in the derivatives of $H_{\mu\nu}$. Hence, collecting all of the
contributions coming from the gravitational part of action
\eqref{acaocompleta}, one gets
\begin{equation}
S^{\ss{(2)}}=\frac{P}{2}\int du\,d^{3}x\frac{1}{u^2}
\left[H_{tx}^{'2}-\mathfrak{h}
H_{yx}^{'2}\right]+...\,,
\label{acaoaxial1}
\end{equation}
where 
\begin{equation}
P=\left(\frac{4\pi T}{3}\right)^{3}\frac{L^{2}}{2\kappa_{4}^2}=
\frac{8\sqrt{2}}{81}\pi^{2}N^{3/2}T^{3}
\label{pressao}
\end{equation}
is interpreted as the pressure of the dual plasma \cite{herzog3}.

Now expressing functions $H_{tx}^{'}$ and $H_{yx}^{'}$ in terms of the axial
fundamental variable $Z_{1}$ through Eqs. \eqref{axial1}-\eqref{mestreaxial},
substituting the resulting relations into Eq. \eqref{acaoaxial1}, and making
use of the fundamental equation \eqref{yeqaxial}, it is found the following
action (at the boundary)
\begin{equation}
S_{\ss{boundary}}^{\ss{(2)}}=\frac{P}{2}\;\underset{u\rightarrow 0}{
\mbox{lim}}\,\int\frac{d\mathfrak{w}\, d\mathfrak{q}}{(2\pi)^{2}}\,
\frac{\mathfrak{h}}{u^{2}(\mathfrak{w}^{2}-\mathfrak{q}^{2}\mathfrak{h})}
Z_{1}^{'}(u,k)Z_{1}(u,-k)+ S_{\ss{CT}}^{\ss{(2)}},
\label{axial-acao2}
\end{equation}
where the contact terms represented by $S_{\ss{CT}}^{\ss{(2)}}$ do not carry
derivatives of the metric perturbation functions. In the calculation of the
correlation functions, after Fourier transformation, the contact terms give
rise to derivatives of the Dirac delta function. Their removal can be done
through the holographic renormalization, with the inclusion of appropriate
counter terms in the supergravity action \cite{bianchi}.

Besides using Eq. \eqref{mestreaxial}, the asymptotic expansion given by Eq.
\eqref{assintgrav1} is used to write the derivative of the gauge-invariant
quantity $Z_{1}$ in terms of the boundary values of the perturbation fields 
$H_{\mu\nu}^{0}(k) =H_{\mu\nu}(u\rightarrow 0,k)$, and then the AdS/CFT
prescription \cite{son1} can be applied to the present case in order to
calculate the retarded correlation functions of the holographic stress-energy
tensor $T^{\mu\nu}$. The result is
\begin{align}
&G_{tx,tx}=-3P\frac{\mathfrak{q}^{2}}{
(\mathfrak{w}^{2}-\mathfrak{q}^{2})}\frac{\mathcal{B}_{(1)}
(\mathfrak{w},\mathfrak{q})}{\mathcal{A}_{(1)}
(\mathfrak{w},\mathfrak{q})},
\label{correlacao-tftf}\\
&G_{tx,yx}=3P\frac{\mathfrak{w}\mathfrak{q}}{
(\mathfrak{w}^{2}-\mathfrak{q}^{2})}\frac{\mathcal{B}_{(1)}
(\mathfrak{w},\mathfrak{q})}{\mathcal{A}_{(1)}
(\mathfrak{w},\mathfrak{q})},
\label{correlacao-tfyf}\\
&G_{yx,yx}=-3P\frac{\mathfrak{w}^{2}}{
(\mathfrak{w}^{2}-\mathfrak{q}^{2})}\frac{\mathcal{B}_{(1)}
(\mathfrak{w},\mathfrak{q})}{\mathcal{A}_{(1)}
(\mathfrak{w},\mathfrak{q})}.
\label{correlacao-yfyf}
\end{align}

As it happens for the current-current correlations functions, one can
find general expressions for the two-point thermal functions associated
to the stress-energy tensor which hold for any scale invariant
$(2+1)$-dimensional field theory (see Appendix \ref{apen-correlations}).
 For fluctuations of the transverse momentum density in the CFT
one has the correlators
\begin{align}
&G_{tx,tx}= \frac{\mathfrak{q}^{2}} {2\left(\mathfrak{w}^2
-\mathfrak{q}^{2}\right)} \,G_{1}(\mathfrak{w},\mathfrak{q}),
                        \label{emtcorrfun1}\\
&G_{tx,yx}= -\frac{\mathfrak{w}\mathfrak{q}}{2\left(\mathfrak{w}^2
-\mathfrak{q}^{2}\right)}\,G_{1}(\mathfrak{w},\mathfrak{q}),
                       \label{emtcorrfun2}\\
&G_{yx,yx}=\frac{\mathfrak{w}^{2}}{2\left(\mathfrak{w}^2
-\mathfrak{q}^{2}\right)} \,G_{1}(\mathfrak{w},\mathfrak{q}). 
                      \label{emtcorrfun3}
\end{align}
Therefore, by comparing the general expressions of Eqs.
\eqref{emtcorrfun1}-\eqref{emtcorrfun3} to the results given in Eqs.
\eqref{correlacao-tftf}-\eqref{correlacao-yfyf}, the following scalar
function is found 
\begin{equation}
G_{1}(\mathfrak{w},\mathfrak{q})=-6P\,\frac{\mathcal{B}_{(1)}
(\mathfrak{w},\mathfrak{q})}{\mathcal{A}_{(1)}(\mathfrak{w},
\mathfrak{q})}.\label{G_1}
\end{equation}
It is then seen that Dirichlet condition imposed on the fundamental variable
$Z_{1}$ at the boundary, $Z_{1}(0)=\mathcal{A}_{(1)}(\mathfrak{w},
\mathfrak{q})=0,$ leads straightforwardly to the poles of the correlation
functions $G_{tx,tx}$, $G_{tx,yx}$ and  $G_{yx,yx}$. As a consequence of this
result, such a requirement also yields the quasinormal spectrum associated to
the axial gravitational perturbation modes of plane-symmetric black holes.

\subsubsection{Polar sector}

The procedure to be applied to the polar sector of gravitational
perturbations is the same as for the axial sector. The starting point here is
the part of the boundary gravitational action built with the quadratic terms
in the polar metric perturbations, which is given by \cite{herzog3}
\begin{equation}
\begin{split}
S_{\ss{boundary}}^{\ss{(2)}}=\frac{P}{2}\;
\underset{u\rightarrow 0}{\mbox{lim}}\,\int d^{3}x\bigg[&
\frac{1}{4}\left(2H_{tt}^{2}-8H_{ty}^{2}+H_{tt}H_{xx}+
H_{tt}H_{yy}\right)-\frac{1}{4}\left(H_{xx}-H_{yy}\right)^{2}\\
&-\frac{\mathfrak{h}}{2u^{2}}\left(H_{ty}^{2}+
H_{xx}H_{yy}-H_{tt}H_{xx}
-H_{tt}H_{yy}\right)^{'}\bigg].
\end{split}
\label{polar-acao1}
\end{equation}
By using the relation among polar metric fluctuations and the gauge-invariant
variable $Z_{2}$, Eq. \eqref{mestrepolar}, and the equations of motion
\eqref{polar1}-\eqref{polar4}, the boundary action \eqref{polar-acao1} can be
cast into the form
\begin{equation}
S_{\ss{boundary}}^{\ss{(2)}}=\frac{P}{2}\;\underset{u\rightarrow
0}{\mbox{lim}}
\,\int\frac{d\mathfrak{w}d\mathfrak{q}}{(2\pi)^{2}}\,
\frac{\mathfrak{h}}{u^{2}\left[4\mathfrak{w}^{2}-
\mathfrak{q}^{2}(4-u^{3})\right]^{2}}\,Z_{2}^{'}(u,k)
Z_{2}(u,-k)+S_{\ss{CT}}^{\ss{(2)}},
\label{polar-acao2}
\end{equation}
where the contact terms $S_{\ss{CT}}^{\ss{(2)}}$ do not contain derivatives
of metric perturbations. One now uses the asymptotic expansion
\eqref{assintgrav2} to write the derivative of the gauge-invariant variable
$Z_2$ in terms of boundary values of the polar metric perturbations
$H_{\mu\nu}^{0}(k)$. After substituting the resulting expression into the
action \eqref{polar-acao2}, the appropriate functional
derivatives\footnote{Functional derivatives in the sense defined by the
Lorentzian AdS/CFT prescription of Ref. \cite{son1}.} of the action with
respect to
the independent fields $H_{tt}^{0}(k)$, $H_{ty}^{0}(k)$, $H_{yy}^{0}(k)$ and
$H_{xx}^{0}(k)$ are performed. Notice, however, that in the case of polar
gravitational perturbations, the use of the AdS/CFT prescription is not
direct. In fact, it is first necessary to identify explicitly how the metric
perturbations couple to the stress-energy tensor at the boundary. As
discussed in Refs. \cite{liu, policastro2}, such a coupling
is given by
\begin{equation}
-\frac{1}{2}\int dt\, d^{2}x\,h_{\;\,\mu}^{\nu}T_{\;\;\nu}^{\mu}=
-\frac{1}{2}\int\,dt\,d^{2}x\left[H_{tt}^0T^{tt}
+H_{xx}^{0}T^{xx} +H_{yy}^{0}T^{yy}+2H_{ty}^{0}T^{ty}\right].
\end{equation}
Taking this coupling into account, the covariant components of the polar
correlation functions are found 
\begin{equation}
G_{\mu\nu,\alpha\beta}=Q_{\mu\nu,\alpha\beta}
G_{2}(\mathfrak{w},\mathfrak{q}),
\end{equation}
where the scalar function $G_{2}(\mathfrak{w},\mathfrak{q})$
is given by
\begin{equation}
G_{2}(\mathfrak{w},\mathfrak{q})=
-6P\frac{\mathcal{B}_{(2)}(\mathfrak{w},\mathfrak{q})}{
\mathcal{A}_{(2)}(\mathfrak{w},\mathfrak{q})}+\mbox{contact terms},
\end{equation}
and tensor $Q_{\mu\nu,\alpha\beta}$ is given in Appendix
\ref{apen-correlations} (see also \cite{kovtun1}). This result shows
definitely that Dirichlet boundary condition imposed on the fundamental
variable $Z_{2}$ at infinity leads to the poles of the function
$G_{2}(\mathfrak{w},\mathfrak{q})$, and, by definition, to the quasinormal
frequencies of polar gravitational vibration modes.

\subsection{QNM and the gauge-invariant variables}

As in the study of electromagnetic quasinormal modes, a comparison between
results found using RWZ variables with the ones obtained using
KS gauge-invariant variables to describe gravitational
perturbations deserves to be made. As in that case, while the axial
quasinormal spectrum is independent of the choice of the fundamental
variable, polar quasinormal spectrum strongly depends on it.

A strong evidence that the QNM spectra obtained using either RWZ or KS
variables to describe axial gravitational perturbations are identical
comes from the study of the hydrodynamic limit of such perturbations as
performed in Ref. \cite{miranda1} and in Sect. \ref{hydro-grav} (see below).
As a matter of fact, even though different methods have been employed in
each case, both of the quasinormal spectra present a typical hydrodynamic
shear mode with diffusion coefficient  $D=1/4\pi T$, independently if one
uses variable $Z^{\ss{(-)}}$, or variable $Z_{1}$. Furthermore,  it can be
shown that the explicit relation between RWZ and KS variables is 
\begin{equation}
Z_{1}=\frac{f}{r^{2}}\partial_{r} \left[rZ^{\ss{(-)}}\right],
\end{equation}
and at the AdS spacetime boundary, axial fundamental variables $Z^{\ss{(-)}}$
and  $Z_{1}$ are proportional to each other,
\begin{equation}
Z_{1}(u)\big|_{u=0}=\frac{1}{L^{2}}Z^{\ss{(-)}}(u)\big|_{u=0}\, ,
\end{equation}
what proves that the two spectra obtained from $Z^{\ss{(-)}}$ and $Z_{1}$
are indeed identical to each other.

The situation in the polar sector is quite diverse from what happens in the
axial sector. As it is shown in the sequence of the present work (see Sects.
\ref{hydro-grav} and \ref{dispersion-gravit}), RWZ and KS variables with the
same boundary conditions generate different quasinormal frequencies. In
particular, it is shown in Sect. \ref{hydro-grav} that the hydrodynamic limit
of $Z_{2}$ contains a sound wave mode which is not seen in the quasinormal
spectrum obtained from $Z^{\ss{(+)}}$.

\subsection{Dispersion relations for the hydrodynamic QNM}
\label{hydro-grav}

For the fluctuations of the stress-energy tensor in the dual field theory,
hydrodynamics predicts a shear mode in the transverse (axial) sector and a
sound wave mode in the longitudinal (polar) sector. As it is going to be
shown below, these modes also appear in the gravitational QNM spectra of the
plane black holes as long as one investigates  the regime of small
frequencies and wavenumbers, $\mathfrak{w}\rightarrow 0$ and
$\mathfrak{q}\rightarrow 0$. Proceeding analogously to the case of
electromagnetic perturbations studied in Sect. \ref{hydro-eletro}, the
following change of variables is done
\begin{equation}\label{Z312}
H_{j}(u)=\mathfrak{h}^{i\mathfrak{w}/3}Z_{j}(u),\quad\qquad j=1,2.
\end{equation}
Functions $H_{1}$ and $H_{2}$ are then expanded in power series of
$\mathfrak{w}$ and $\mathfrak{q}$. Besides being approximate solutions to
Eqs. \eqref{yeqaxial} and \eqref{yeqpolar} in the hydrodynamic limit, such
series must represent ingoing waves at the horizon, namely,
$H_{j}(u)\big|_{u=1}= {\rm constant}$. These conditions are fulfilled by the
following expansions:
\begin{equation}
Z_{1}=C_{1}\mathfrak{h}^{-i\mathfrak{w}/3}
\left[1+i\frac{\mathfrak{q}^{2}\mathfrak{h}}{3\mathfrak{w}}
+{\mathcal{O}}(\mathfrak{w}^{2})\right],\\
\end{equation}
\begin{equation}
Z_{2}=C_{2}\mathfrak{h}^{-i\mathfrak{w}/3}\left[2-u^{6}
-4\frac{\mathfrak{w}^{2}}{\mathfrak{q}^{2}}-
\frac{4i\mathfrak{w}\mathfrak{h}}{3}+{\mathcal{O}}(\mathfrak{w}^{2})
\right],\\
\end{equation}
where $C_{1}$ and $C_{2}$ are arbitrary normalization constants.  Dirichlet
conditions at the spacetime boundary, $u=0$, are now imposed on both
the axial and the polar fundamental variables, $Z_{1}(0)=0$ and $Z_{2}(0)=0$.
The first variable leads to an axial quasinormal mode identified to the shear
mode with
\begin{equation}
\omega=-iDq^{2},
\label{shear}
\end{equation}
while the second variable furnishes a dispersion relation 
characteristic to a sound wave mode
\begin{equation}
\omega=\pm\frac{1}{\sqrt{2}}q-\frac{iDq^{2}}{2}\, ,
\label{soundwave}
\end{equation}
where $D$ is the diffusion coefficient, given by  
\begin{equation}
D=\frac{\eta}{\varepsilon+P}=\frac{1}{4\pi T}, \label{Dcoeff}
\end{equation}
with $\eta$ and $\varepsilon$ being respectively the shear coefficient and the
energy density of the dual system.
\FIGURE{
%\begin{figure}[t]
\centering \epsfig{file=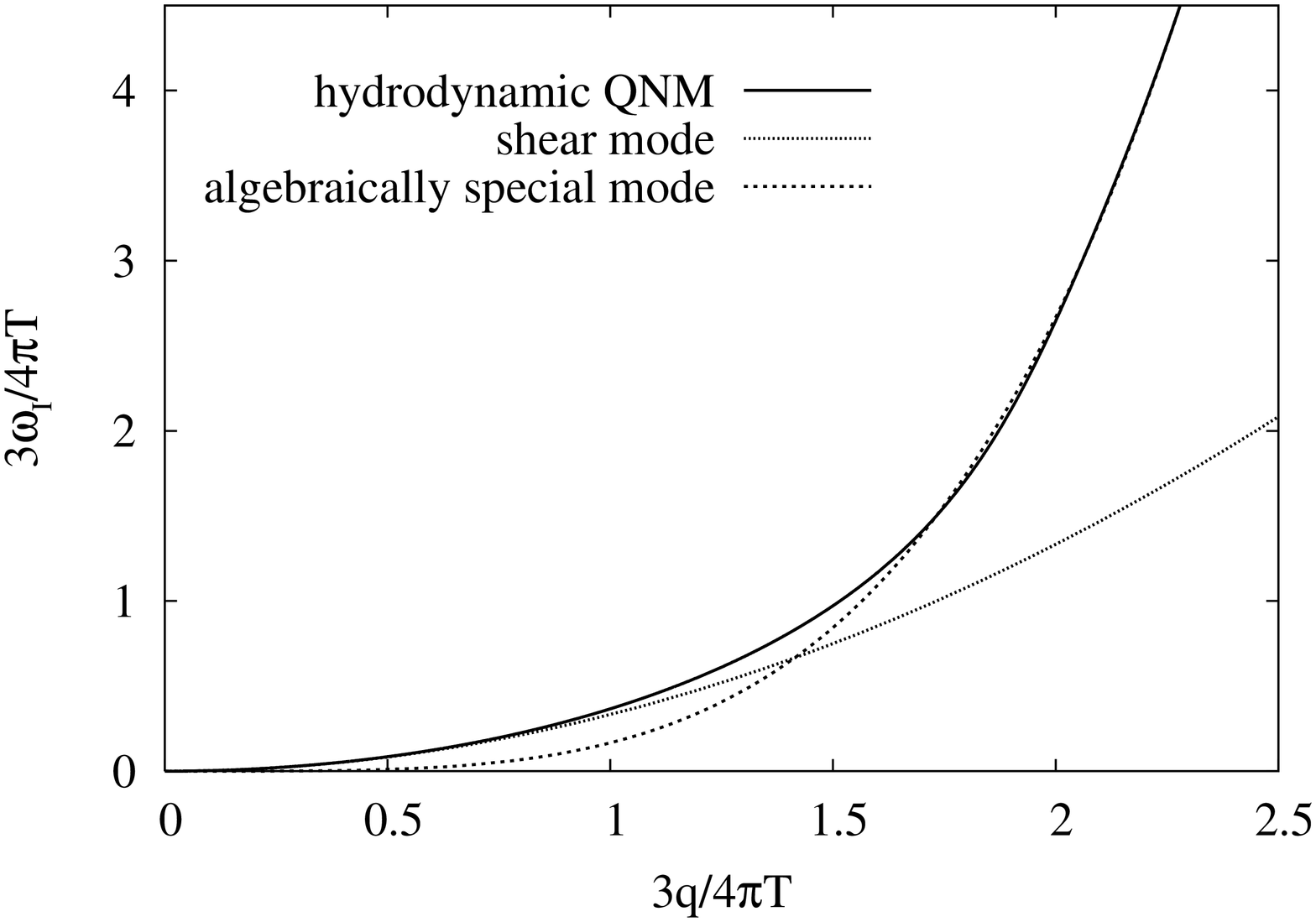, height=9.41cm, width=6.0cm, angle=270}
\caption{The dispersion relation (imaginary part) for the axial
gravitational
hydrodynamic mode (solid line), in comparison to the shear mode,
$\mathfrak{q}^{2}/3$ (dotted line), and to the algebraically special
mode, $\mathfrak{q}^{4}/6$ (dashed line).}
\label{hidrograv_axial}}
%\end{figure}

Among other interesting results that can be found, combining Eq.
\eqref{Dcoeff} to the thermodynamic Euler relation $P=-\varepsilon+Ts\Rightarrow
s=(\varepsilon+P)/T$, one gets the ratio between the shear coefficient and the
entropy density of the dual plasma,
\begin{equation}
\frac{\eta}{s}=\frac{\eta\,T}{\varepsilon+P}=\frac{1}{4\pi}, 
\end{equation}
or, in conventional units, $\eta/s=\hbar/4\pi k_{\ss{B}}$. This ratio is the
same for all finite temperature field theories with a dual gravitational
description in the AdS spacetime \cite{kovtun2}. It is also speculated that
$\eta/s=1/4\pi$ represents a  lower bound --the KSS bound-- of such a ratio
for all fluids in nature.

The complete dispersion relations for the gravitational hydrodynamic
QNM  are obtained by means of the Horowitz-Hubeny method
\cite{horowitz1}, and, for the sake of simplicity, the numerical results for
the axial and polar sectors are analyzed separately.\\

\noindent
{\it (i) Axial modes}: The extension of the dispersion relation of the axial
hydrodynamic QNM for large values of $\mathfrak{q}$ was already done in Ref.
\cite{miranda1}, but for completeness it is also shown in Fig.
\ref{hidrograv_axial}. As the normalized wavenumber $\mathfrak{q}$ reaches
values beyond of the hydrodynamic regime, the magnitude of the QNM frequency
$\mathfrak{w}=-i\mathfrak{w}_{I}$ increases faster than the magnitude of the
shear mode frequency $\mathfrak{w}=-i\mathfrak{q}^{2}/3$, and for very large
wavenumber values, $\mathfrak{q}\gg 1$, the hydrodynamic QNM frequency
approaches the algebraically special frequency
$\mathfrak{w}=-i\mathfrak{q}^{4}/6$.\\

\FIGURE{
%\begin{figure}[t]
\centering\epsfig{file=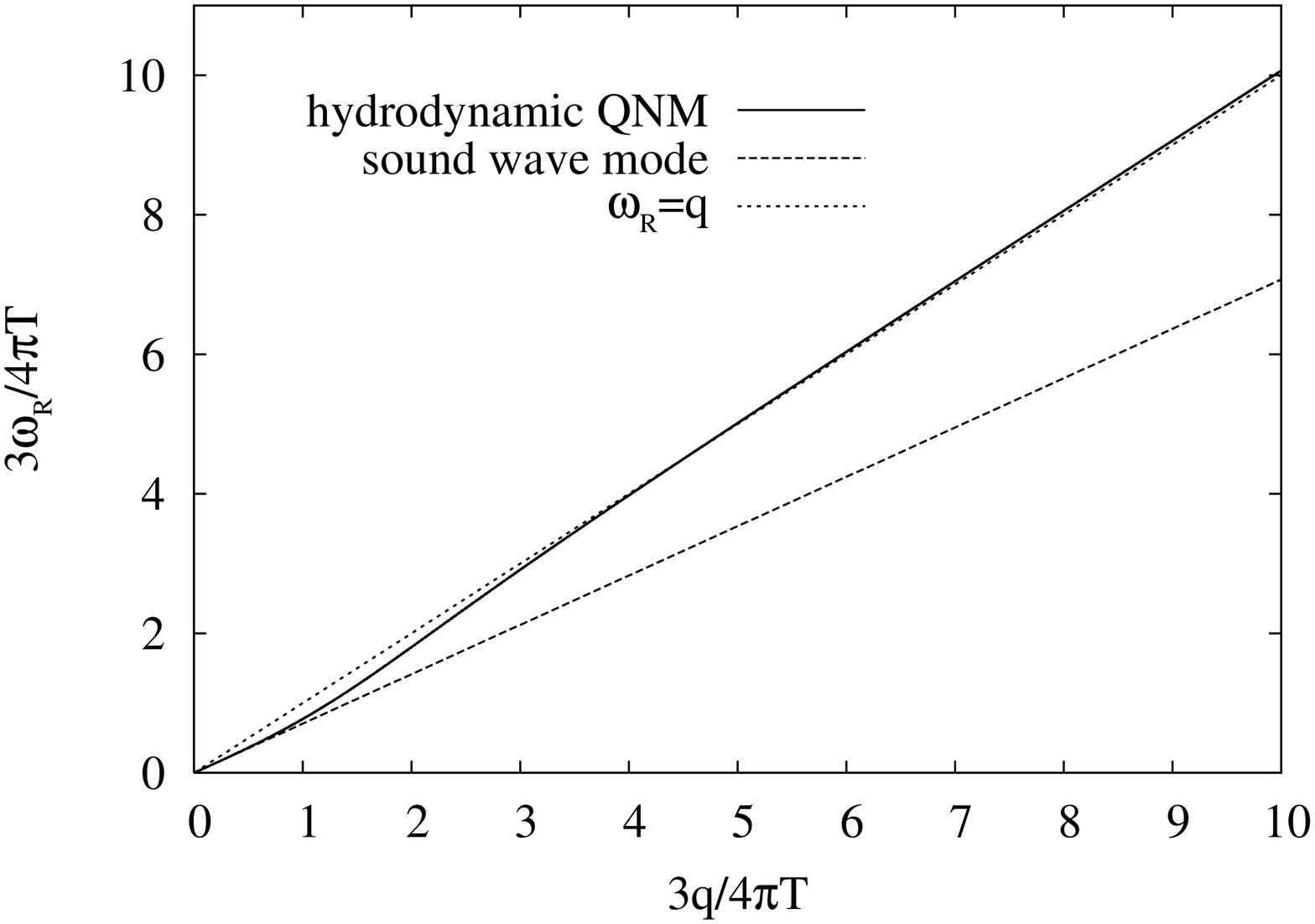, height=7.137cm,
width=4.968cm, angle=270}
\centering\epsfig{file=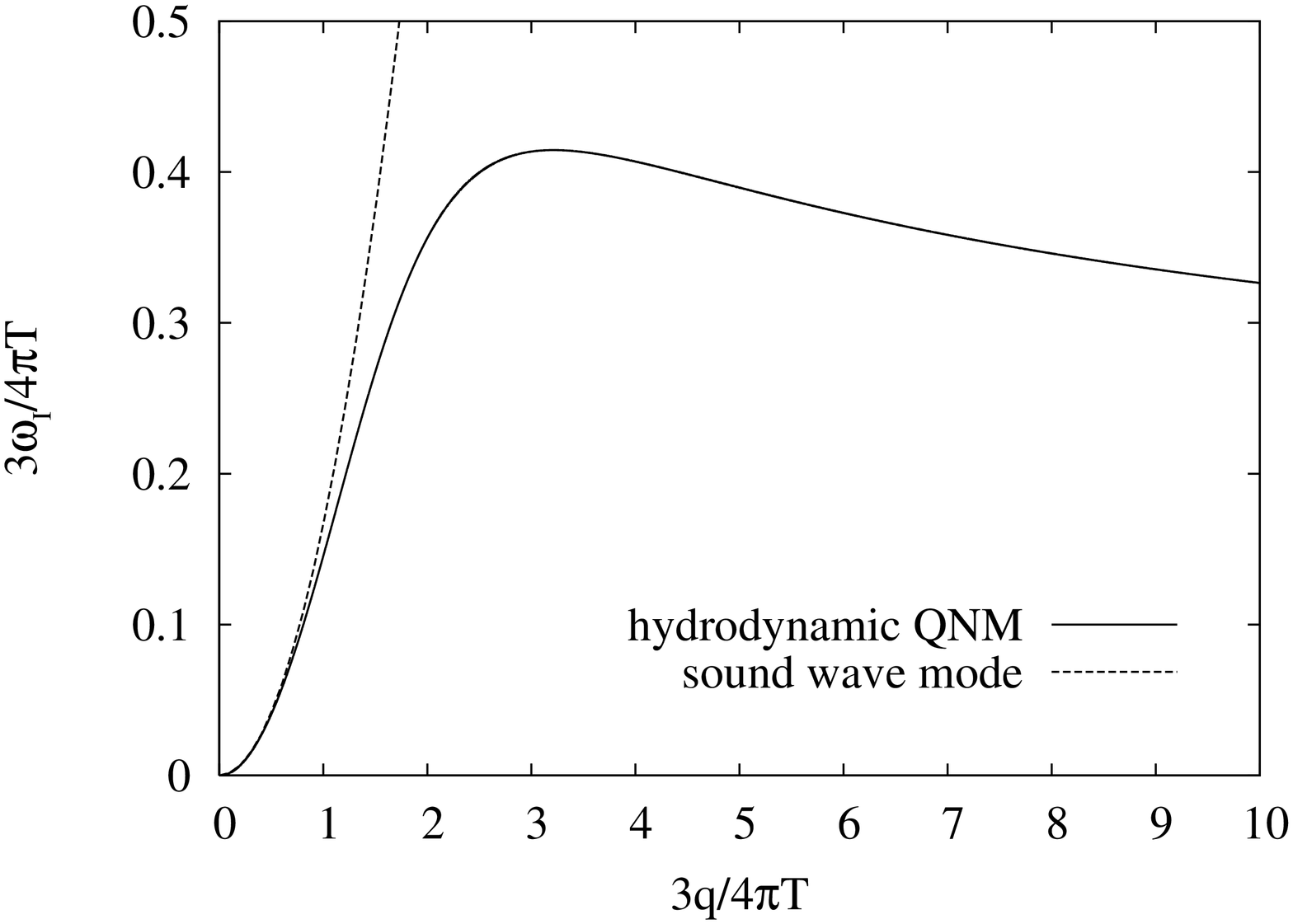, height=7.137cm,
width=4.968cm, angle=270}
\caption{Real and imaginary parts, respectively, of the dispersion
relation $\mathfrak{w}(\mathfrak{q})$ for the hydrodynamic longitudinal
gravitational mode (solid lines). The dashed lines in both of the figures
correspond to the sound wave mode, given by Eq. \eqref{soundwave}, and the
dotted curve in the figure on the left hand side is the relation
$\mathfrak{w}_{R}=\mathfrak{q}$.}
\label{hidrogravpolar}}
%\end{figure}
\FIGURE{
%\begin{figure}[t]
\centering\epsfig{file=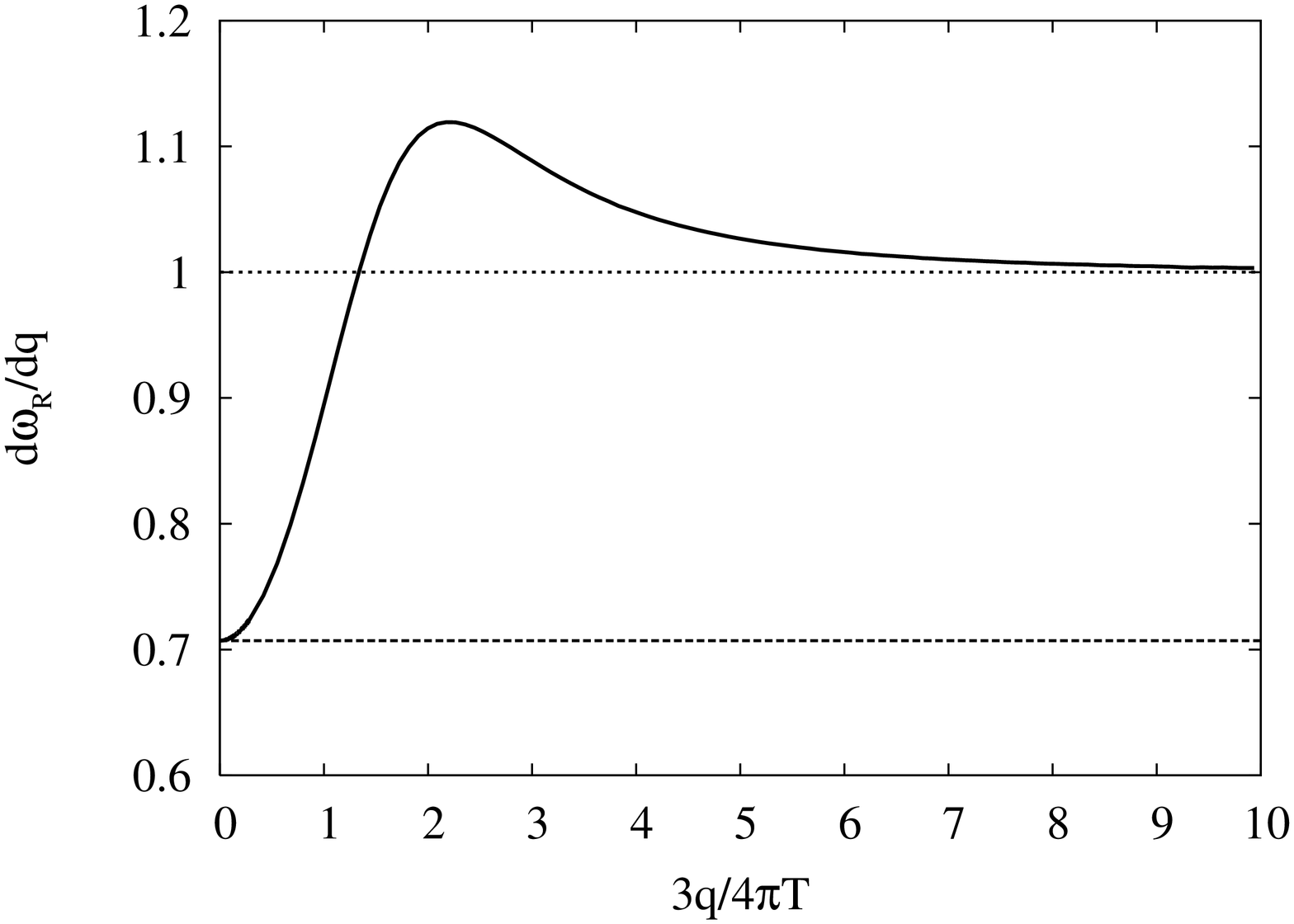, height=7.137cm,
width=4.968cm, angle=270}
\centering\epsfig{file=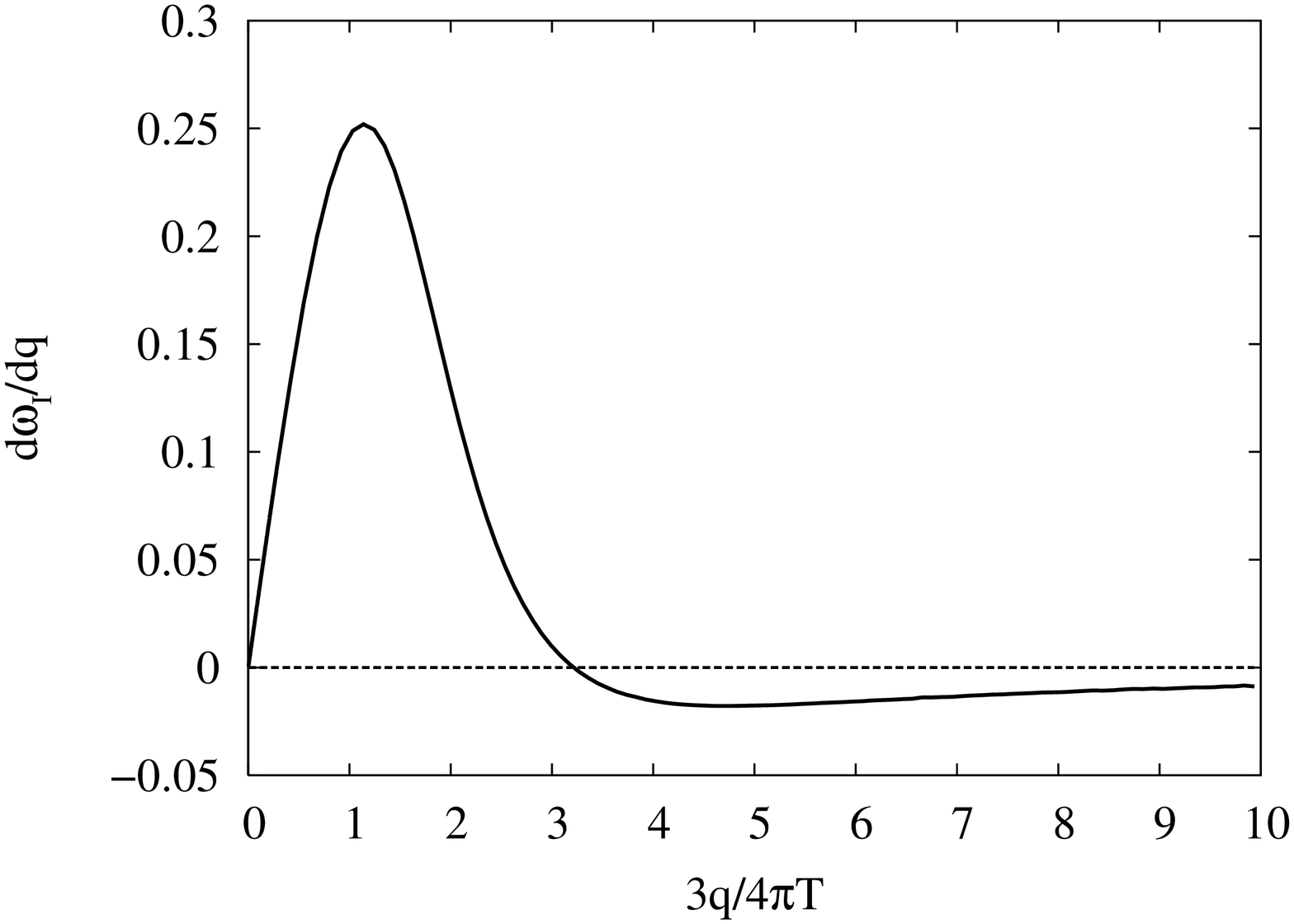, height=7.137cm,
width=4.968cm, angle=270}
\caption{The group velocity $c_{s}=d\mathfrak{w}_{R}/d\mathfrak{q}$ and the
derivative $d\mathfrak{w}_{I}/d\mathfrak{q}$ as a function of the normalized
wavenumber $\mathfrak{q}$  for the polar hydrodynamic gravitational mode
(solid lines).
The dashed line in the figure on the left is the sound velocity for a
(2+1)-dimensional CFT, $c_{s}=1/\sqrt{2}$, and the
dotted line in the same figure represents the speed of light, $c=1$.}
\label{hidroderivada}}
%\end{figure}

\noindent
{\it (ii) Polar modes}: In this sector of perturbations, RWZ and KS variables
submitted to the same boundary conditions generate different quasinormal
frequencies. As seen above, the hydrodynamic limit of $Z_{2}$ presents a
sound wave mode which is not present in the quasinormal spectrum obtained
from $Z^{\ss{(+)}}$. The extended dispersion relations for the longitudinal
hydrodynamic QNM are shown in Fig. \ref{hidrogravpolar}. The real part of the
frequency clearly shows the transition from the hydrodynamic regime
$\mathfrak{w}_{R}= \mathfrak{q}/\sqrt{2}$, at low wavenumbers, to a regime
characterized by collisionless dual plasma in which the dispersion relation
$\mathfrak{w}_{R}(\mathfrak{q})$ approaches the ultra-relativistic relation
$\mathfrak{w}_{R}=\mathfrak{q}$. Between these two extreme regimes, the group
velocity, defined by $c_{s}=d\mathfrak{w}_{R}/d\mathfrak{q}$, assumes values
that are higher than the speed of light. The graphs in Fig.
\ref{hidroderivada} show that $c_{s}>1$ for all wavenumbers larger than
$\mathfrak{q}\simeq 1.336$, and that the minimum decaying time (the maximum
of $\mathfrak{w}_{I}$) corresponds to $\mathfrak{q}\simeq 3.213$. Notice,
however, that $d\mathfrak{w}_{R}/d\mathfrak{q}$ surpasses the speed of light
at wavenumber values lying outside the hydrodynamic regime and therefore that
superluminal group velocity cannot be interpreted as the sound velocity in
the corresponding media.

\subsection{Dispersion relations for the non-hydrodynamic QNM}
\label{dispersion-gravit}

Conventionally, a non-hydrodynamic QNM is every mode for which the dispersion
relation presents a gap in the limit $\mathfrak{q}\rightarrow 0 $. That is to
say, the quasinormal frequency $\mathfrak{w}(\mathfrak{q})$ tends to a
nonzero value in the limit where the wavenumber goes to zero. These kind of
gravitational perturbation modes are studied in this section.

As done in the case of electromagnetic perturbations, the QNM obtained by
using RWZ gravitational variables $Z^{\ss{(\pm)}}$, obeying
Schr\"odinger-like equations, are compared to the QNM obtained by using the
KS gauge-invariant quantities $Z_{1,2}$, which lead to the
poles of stress-energy tensor correlators in the $\mathcal{N}=8$
super-Yang-Mills field theory. In particular, the results obtained here from
$Z_{1,2}$ are compared to the results of Refs. \cite{cardoso5,miranda1}.

\subsubsection{Purely damped modes}

The spectra of gravitational perturbations of plane-symmetric $AdS_{4}$
black holes do not present non-hydrodynamic quasinormal frequencies
with vanishing real part. In fact, the only purely damped mode
of the gravitational perturbations is the axial hydrodynamic QNM
which was already investigated in the last section.

\subsubsection{Ordinary quasinormal modes}
\label{nonhydro-gravit}

As usual, the study of the dispersion relations of regular non-hydrodynamic
gravitational QNM is more conveniently performed by considering axial and
polar sectors of such perturbations separately.\\

\noindent
{\it (i) Axial modes}: As shown above, in the case of gravitational axial
modes both of the fundamental variables $Z_{1}$ and $Z^{\ss{(-)}}$ yield the
same QNM spectrum. Even though a detailed study of the fundamental
non-hydrodynamic QNM (based on the RWZ master variable) was performed in Ref.
\cite{miranda1}, higher overtones of axial gravitational QNM were not fully
investigated. Hence, the aim here is to complete the analysis by including
such higher overtones. Fig. \ref{gravaxial} shows the numerical results for
the dispersion relations of the first five axial quasinormal modes:
$n=1$,..., $5$. The general forms of the curves
are approximately the same for all values of $n$:
At intermediate values of $\mathfrak{q}$, there is a local minimum in the
real part of the frequency $\mathfrak{w}_{R}(\mathfrak{q})$, and for large
values of the wavenumber, every dispersion relation
$\mathfrak{w}_{R}(\mathfrak{q})$ tends to some straight line parallel to the
ultra-relativistic energy-momentum relation,
$\mathfrak{w}_{R}=\mathfrak{q}$.\\

\FIGURE{
%\begin{figure}[t]
\centering\epsfig{file=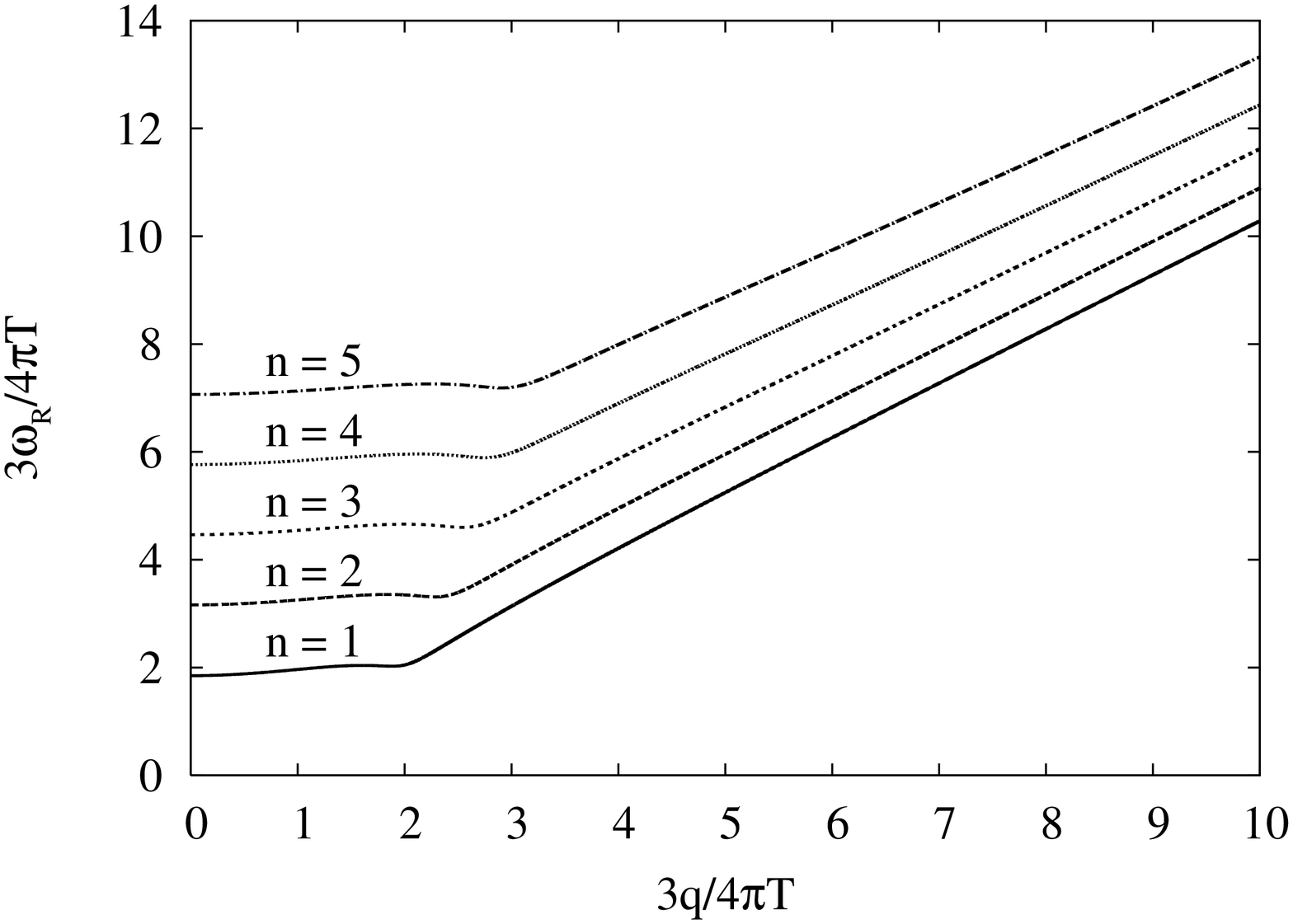, height=7.137cm,
width=4.968cm, angle=270}
\centering\epsfig{file=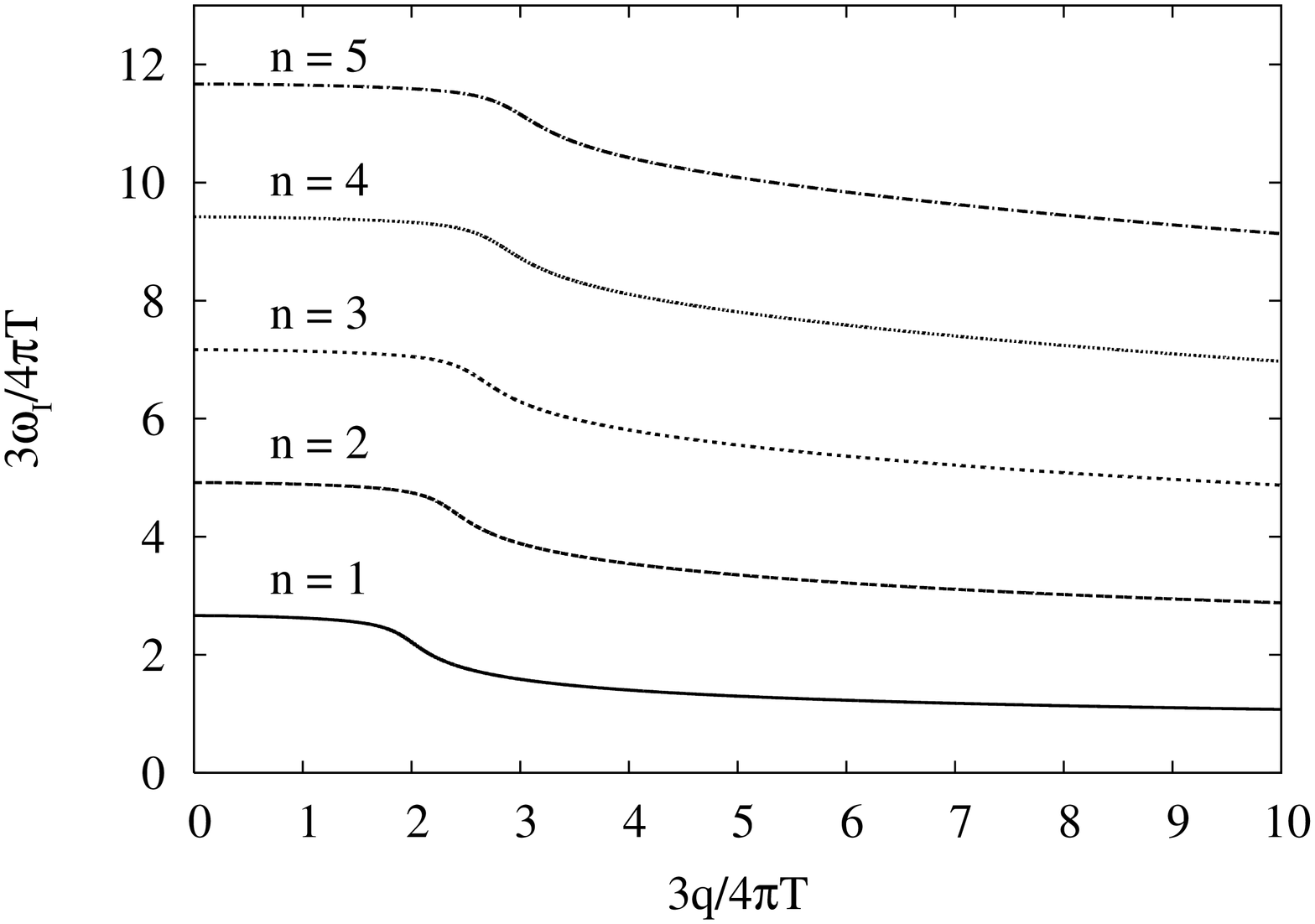, height=7.137cm,
width=4.968cm, angle=270}
\caption{The first five quasinormal frequencies of non-hydrodynamic axial
gravitational modes, $\mathfrak{w}=(3\omega_{R}/4\pi T)-i(3\omega_{I}/4\pi
T)$, as a function of the normalized wavenumber $\mathfrak{q}=3q/4\pi T$.
The quantum number $n$ arranges the modes in growing order according to
the strength of the imaginary parts of the frequencies.}
\label{gravaxial}}
%\end{figure}
\FIGURE{
%\begin{figure}[t]
\centering\epsfig{file=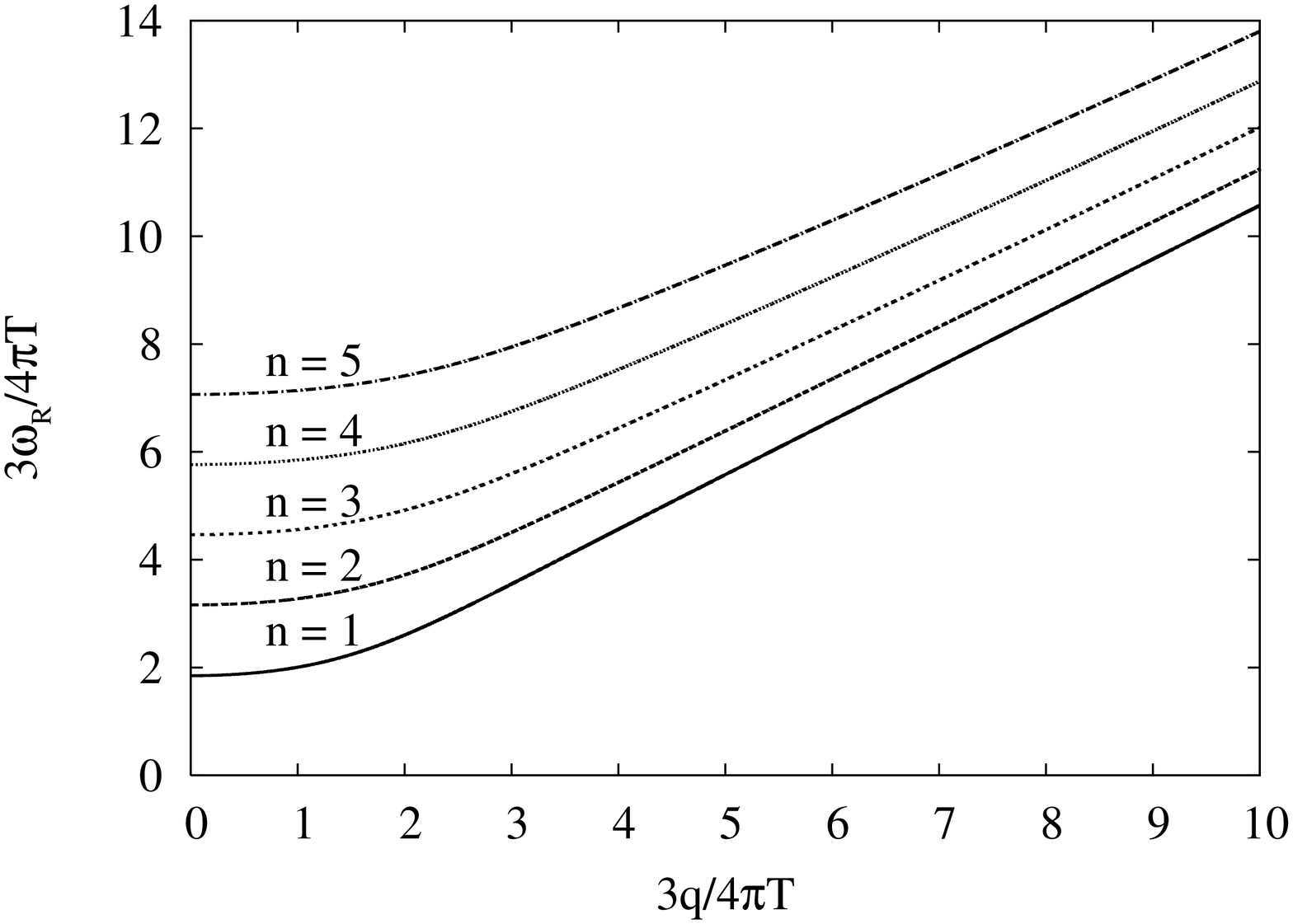, height=7.137cm,
width=4.968cm, angle=270}
\centering\epsfig{file=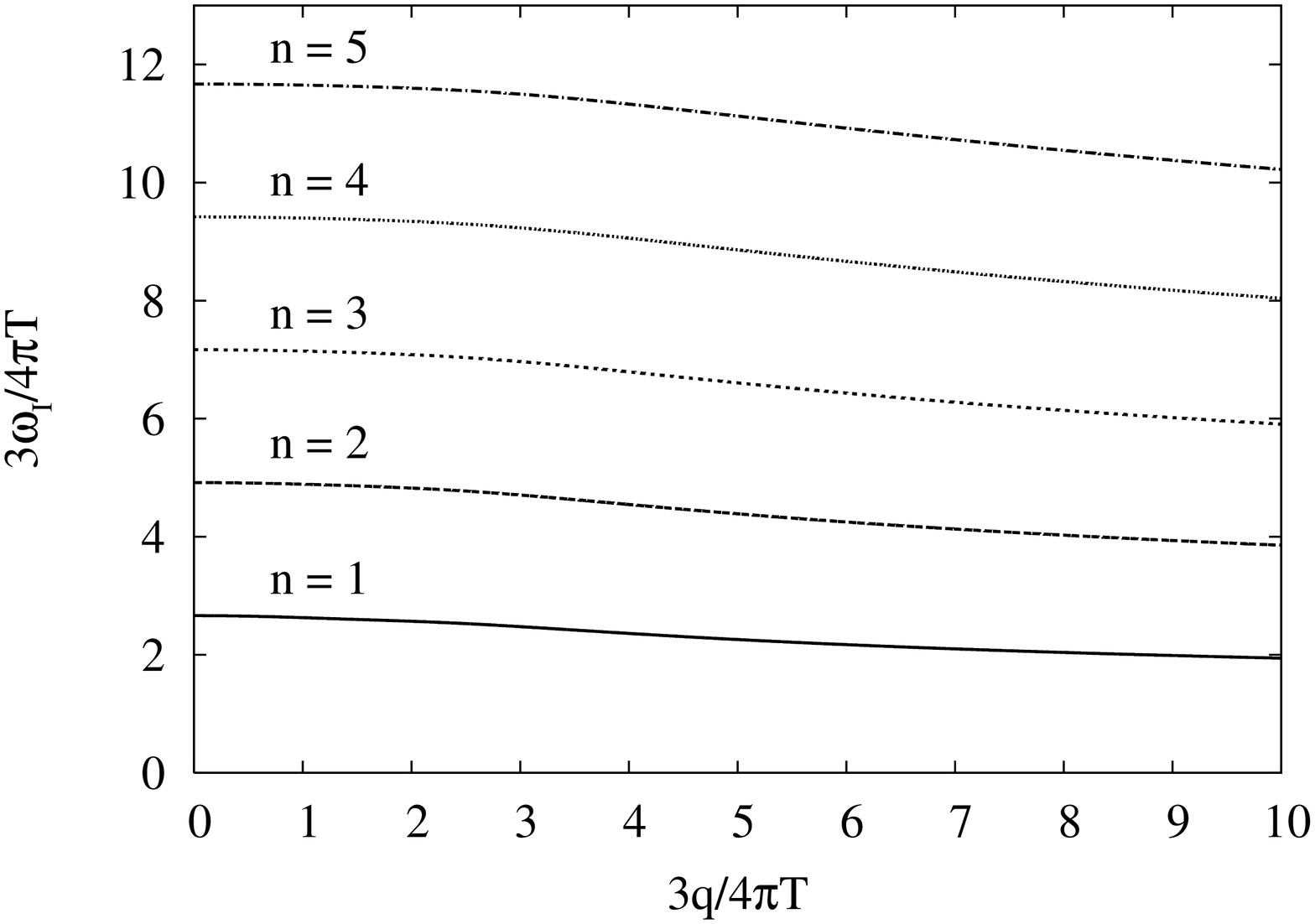, height=7.137cm,
width=4.968cm, angle=270}
\caption{The first five quasinormal frequencies of non-hydrodynamic
polar gravitational modes, $\mathfrak{w}=(3\omega_{R}/4\pi T)
-i(3\omega_{I}/4\pi T)$, as a function of the normalized
wavenumber $\mathfrak{q}=3q/4\pi T$, ordered as in
the case of axial gravitational modes.}
\label{gravpolar} }
%\end{figure}

\noindent
{\it (ii) Polar modes}:
Contrary to the axial perturbations, the longitudinal gravitational
fluctuations present different spectra as one takes $Z_{2}$ or $Z^{\ss{(+)}}$
as fundamental variable. The numerical results for these modes, based on the
gauge-invariant variable $Z_{2}$, are shown in Fig. \ref{gravpolar}. Notice
that now the real part of the quasinormal frequency $\mathfrak{w}_{R}$ is a
monotonic increasing function of $\mathfrak{q}$, showing neither local maxima
nor local minima which, on contrary, appear in the quasinormal spectrum for
the RWZ master variable $Z^{\ss{(+)}}$ used in Refs.
\cite{cardoso5,miranda1}. A comparison between the QNM spectrum found from
the RWZ variable (cf. Ref. \cite{miranda1}) and that found from the
KS variable used here can also be done through the data
presented in Table \ref{tabelapolar}. It is clearly seen the similarity
between the two spectra in the two asymptotic regions of wavenumber values.
Both of the spectra are approximately the same for small values and also for
large values of $\mathfrak{q}$. The main differences happen in the regime
where the normalized wavenumber $\mathfrak{q}$ is of the order of unity,
which means that $q$ is of the order of the blackhole Hawking temperature.
Moreover, the real parts of the quasinormal frequencies in both of the
spectra are relatively closer to each other when compared to the
corresponding imaginary parts. This implies that, even though the QNM
oscillation frequencies are essentially the same for both choices of
variables, the decaying timescales ($\tau=1/\omega_{I}$) are significantly
smaller for the KS choice.
\TABLE{
%\begin{table}[t]
\begin{tabular}{lccccc}
\hline\hline
& \multicolumn{2}{c}{Kovtun-Starinets} & \multicolumn{2}{c}
{Regge-Wheeler-Zerilli}\\
\cline{2-5}
$\quad\mathfrak{q}\quad $ & $\qquad\mathfrak{w}_{R}\qquad $ &
$\qquad\mathfrak{w}_{I}\qquad $
& $\qquad\mathfrak{w}_{R}\qquad $ &
$\qquad\mathfrak{w}_{I}\qquad $\\
\hline
0.004 & 1.84942 & 2.66385 & 1.84945 & 2.66384\\
0.04  & 1.84964 & 2.66379 & 1.85027 & 2.66248\\
0.4  & 1.87207 & 2.65770 & 1.92488 & 2.52658\\
1   & 2.00603 & 2.62917 & 2.03016 & 1.92213\\
2   & 2.60256 & 2.56803 & 2.30526 & 1.55218\\
5   & 5.57791 & 2.25854 & 5.25618 & 1.27974\\
10   & 10.5703 & 1.94304 & 10.2839 & 1.07342\\
\hline\hline
\end{tabular}
\centering
\caption{Numerical results for the first non-hydrodynamic quasinormal mode
associated to the polar gravitational perturbations. The second and third
columns show respectively the values of real and imaginary parts of the
frequencies obtained by using the KS variable $Z_{2}$, and the last two
columns present the results obtained by using the RWZ variable
$Z^{\ss{(+)}}$, which are also shown in Ref. \cite{miranda1}.}
\label{tabelapolar}}
%\end{table}

\section{Final comments and conclusion}
\label{consid-final}

One of the important issues dealt with in the present work is related to the
choice of appropriate variables in order to determine the QNM spectra of AdS
black holes. It is argued that Kovtun-Starinets gauge-invariant quantities,
together with incoming-wave condition at horizon and Dirichlet boundary
condition at infinity, should be used in order to find the correct
quasinormal dispersion relations. The resulting spectrum for a given
perturbation is in general different from what is obtained using other
kind of variables with the same boundary conditions.

In the case of electromagnetic perturbations, contrary to what is obtained
using the so-called Regge-Wheeler-Zerilli quantities $\Psi^{\ss{(\pm)}}$ where
the spectra for both the axial and the polar modes are the same,
Kovtun-Starinets variables present a different spectrum for each mode type.
Also, both these sectors of electromagnetic perturbations present purely
damped modes whose dispersion relations, in the limit of small wavenumbers,
approach the bosonic Matsubara frequencies $\omega=-2i\pi Tn_{s}$. This
result can be compared to the ``quasi-Matsubara'' frequencies, $\omega=2\pi
Tn_{s}(1-i)$, that have been found for zero wavenumber fluctuations of
$(4+1)$-dimensional black branes \cite{nunez}. The studies show the emergence
of infinite sequences of bosonic Matsubara frequencies for both $(3+1)$- and
$(4+1)$-dimensional black branes, but the particular behavior of the QNM
dispersion relations at zero wavenumber is quite different in each case. In
$(3+1)$ dimensions the real part of the frequencies is zero for very small
wavenumber values, while in $(4+1)$ dimensions real and imaginary parts of the
frequencies are both finite at zero wavenumber. Moreover, as pointed out in
Sect. \ref{dispersion-eletro} and discussed in detail in Ref. \cite{herzog4},
the invariance of Maxwell equations under the electric
field $\leftrightarrow$ magnetic field duality operation in
$(3+1)$-dimensional spacetimes implies that there are no electromagnetic QNM
at zero wavenumber. Such a duality invariance does not hold in higher
dimensional spacetimes, what justifies the different behavior of zero
wavenumber electromagnetic QNM found here when compared to the results of Ref.
\cite{nunez}.

Other special property of electromagnetic fluctuations deserving to be
mentioned here is the cutoff in the dispersion relations of
purely damped QNM at a particular value of the wavenumber,
$\mathfrak{q}=\mathfrak{q}_{\mbox{\scriptsize{lim}}}$. This cutoff implies in
an abrupt change in the behavior of the fundamental quasinormal mode, but
not in the thermalization time $\tau$. For wavenumbers in the interval
$0<\mathfrak{q}<\mathfrak{q}_{\mbox{\scriptsize{lim}}}$, parameter $\tau$ is
given by the first purely damped mode, $\tau=1/\omega_{s}$, while for
wavenumber values above $\mathfrak{q}_{\mbox{\scriptsize{lim}}}$, the
characteristic decaying time are governed by the fundamental ordinary QNM,
$\tau=1/\omega_{I}$. Since the imaginary parts of the frequencies of these
modes are equal for $\mathfrak{q}=\mathfrak{q}_{\mbox{\scriptsize{lim}}}$, the
thermalization time changes continuously for wavenumbers close to
$\mathfrak{q}_{\mbox{\scriptsize{lim}}}$.

The numerical results for the gravitational QNM show that the thermalization
time of axial modes for wavenumbers in the interval $0<\mathfrak{q}<1.935$ is
determined by the hydrodynamic mode. This means that, at least in the limit
$\mathfrak{q}\ll 1$, where the shear mode is a good approximation for the
QNM, the thermalization time $\tau$ is a linear function of Hawking
temperature, $\tau\simeq 4\pi T/q^{2}$. As in the case of
electromagnetic fluctuations, the transition to the regime where the
thermalization time is determined by the first regular axial QNM is
continuous. Such a transition happens for the values $\mathfrak{q}\simeq
1.935$ and $\mathfrak{w}_{I}\simeq 2.296$. Also, at this point the
thermalization time reaches its minimum value, $\tau\simeq 0.104/T$. On the
other hand, for polar gravitational perturbations the decaying time $\tau$
is always determined by the hydrodynamic QNM which reduces to the sound wave
mode in the small wavenumber limit $\mathfrak{q}\ll 1$.

Finally, the behavior of the group velocity $c_{s}$ shown in Fig.
\ref{hidroderivada}, which is greater than unity for all
$\mathfrak{q}>1.336$, deserves further analysis. First note that apparent
superluminal propagation of this type, which at first sight seems to violate
causality, has been found in other relativistic quantum systems. For
instance, Scharnhorst \cite{schar} has shown in the case of Casimir effect
that when vacuum fluctuations obey periodic boundary conditions, the two-loop
corrections to the polarization tensor lead to superluminal photon
propagation. Also, it is argued that  the physically meaningful propagation
velocity and, consequently, the one that defines the light cones in
spacetime, is the front wave speed $v_{\mbox{\scriptsize{wf}}}$, which is
given by the limit of the phase speed $v_{\mbox{\scriptsize{ph}}}=\omega/q$
when $\omega\rightarrow\infty$ (see, e.g., Ref. \cite{shore} for a review).
The graph on the left in Fig. \ref{hidroderivada} shows that
$v_{\mbox{\scriptsize{ph}}}$ approaches $c=1$ at high frequencies, and,
therefore, the oscillations in the super-Yang-Mills plasma do not violate
causality.

\section*{Acknowledgments}

We thank conversations with Vitor Cardoso, Marc Casals, and Jos\'e
P. S. Lemos. ASM thanks Funda\-\c{c}\~{a}o Universidade Federal do ABC for
hospitality. VTZ thanks Funda\c c\~ao de Amparo \`a
Pesquisa do Estado de S\~ao Paulo (FAPESP) for financial help. VTZ is
supported by a fellowship from Conselho Nacional de Desenvolvimento Cient\'\i
fico e Tecnol\'ogico of Brazil (CNPq).

\appendix
\section{Correlators in a $(2+1)$-dimensional CFT}
\label{apen-correlations}

The Lorentz index structure of retarded Green functions of conserved currents
and stress-energy tensor in a $D$-dimensional relativistic quantum field
theory was discussed in details in Ref. \cite{kovtun1}. It was shown that the
field theory correlators can be expressed in terms of a set of scalar
functions.
For the present work, a particularly interesting example is that of a
$(2+1)$-dimensional finite temperature conformal field theory. In this
specific case, with the wave three-vector in the form
$k_{\mu}=(-\omega,0,q)$, the transverse component of the current-current
correlation functions can be written as
\begin{equation}
C_{x^{1}x^{1}}(k)
=\Pi^{T}(\omega,q) \label{tself-en}.
\end{equation}
The longitudinal components in turn are given by
\begin{align}
&C_{tt}(k)=\frac{q^{2}}{(\omega^{2}-q^{2})}
\Pi^{L}(\omega,q),\\
&C_{tx^{2}}(k)=-\frac{\omega q}{(\omega^{2}-q^{2})}
\Pi^{L}(\omega,q),\\
&C_{x^{2}x^{2}}(k)=\frac{\omega^{2}}{
(\omega^{2}-q^{2})}\Pi^L(\omega,q), \label{lself-en3}
\end{align}
where $\Pi^{T}(\omega,q)$ and $\Pi^L(\omega,q)$ are two independent scalar
functions. All the correlators of transverse momentum density are expressed
in terms of a scalar function $G_{1}(\omega,q)$:
\begin{align}
&G_{tx^{1},tx^{1}}(k)=\frac{1}{2}\frac{q^{2}}{(\omega^{2}-q^{2})}
G_{1}(\omega,q),\\
&G_{tx^{1},x^{1}x^{2}}(k)=-\frac{1}{2}\frac{\omega q}{
(\omega^{2}-q^{2})}G_{1}(\omega,q),\\
&G_{x^{1}x^{2},x^{1}x^{2}}(k)=\frac{1}{2}\frac{\omega^{2}}{
(\omega^{2}-q^{2})}G_{1}(\omega,q).
\end{align}
In a similar way, the correlators  of longitudinal momentum density, energy
density, and diagonal stress are determined by another scalar function
$G_{2}(\omega,q)$,
\begin{equation}
G_{\mu\nu,\alpha\beta}(k)=Q_{\mu\nu,\alpha\beta}(\omega,q)
G_{2}(\omega,q),
\end{equation}
where the components of the projector $Q_{\mu\nu,\alpha\beta}$ are given by: 
\begin{alignat}{3}
& Q_{tt,tt}=\frac{1}{2}\frac{q^{4}}{(\omega^{2}
-q^{2})^{2}},&\qquad\quad
& Q_{tt,tx^{2}}=-\frac{1}{2}\frac{\omega q^{3}}{
(\omega^{2}-q^{2})^{2}},\\
& Q_{tt,x^{1}x^{1}}=-\frac{1}{2}\frac{q^{2}}{(\omega^{2}
-q^{2})},&\qquad\quad
& Q_{tt,x^{2}x^{2}}=\frac{1}{2}\frac{\omega^{2}q^{2}}{(\omega^{2}
-q^{2})^{2}},\\
& Q_{x^{1}x^{1},x^{1}x^{1}}=\frac{1}{2},&\qquad\quad
& Q_{x^{1}x^{1},tx^{2}}=\frac{1}{2}\frac{\omega q}{
(\omega^{2}-q^{2})},\\
& Q_{x^{1}x^{1},x^{2}x^{2}}=-\frac{1}{2}
\frac{\omega^{2}}{(\omega^{2}-q^{2})},&\qquad\quad
& Q_{x^{2}x^{2},x^{2}x^{2}}=\frac{1}{2}
\frac{\omega^{4}}{(\omega^{2}-q^{2})^{2}},\\
& Q_{x^{2}x^{2},tx^{2}}=-\frac{1}{2}
\frac{\omega^{3}q}{(\omega^{2}-q^{2})^{2}},&\qquad\quad
& Q_{tx^{2},tx^{2}}=\frac{1}{2}
\frac{\omega^{2}q^{2}}{(\omega^{2}-q^{2})^{2}}.
\end{alignat}\\

\end{document}